\documentclass[useAMS,usenatbib]{mn2e}

\usepackage{graphicx}

\newcommand{\ha}{H$\alpha$}

\def\vhel{\ifmmode{V_{{\rm HEL}}}\else{$V_{{\rm HEL}}$}\fi}
\def\vsys{\ifmmode{V_{\rm sys}}\else{$V_{\rm sys}$}\fi}
\def\kms{\ifmmode{~{\rm km\,s}^{-1}}\else{~km~s$^{-1}$}\fi}
\def\vlsr{\ifmmode{v_{\rm lsr}}\else{$v_{\rm lsr}$}\fi}

\def\ltsim{\ifmmode\stackrel{<}{_{\sim}}\else$\stackrel{<}{_{\sim}}$\fi}
\def\gtsim{\ifmmode\stackrel{>}{_{\sim}}\else$\stackrel{>}{_{\sim}}$\fi}

\newcommand{\um}{$\mu$m}
\newcommand{\uk}{$\mu$K}
\newcommand{\qrms}{$Q_{rms-PS}$}
\newcommand{\n}{$n$}

\newcommand{\cnu}{${\mathbf c}_{\nu}$}
\newcommand{\te}{${\mathrm T_e}$}
\renewcommand{\fd}{${f_d}$}

\begin{document}

\title[Reappraising the $COBE$-DMR data]{Reappraising foreground
contamination in the $COBE$-DMR data}

\author[A. J. Banday et
al.]{A. J. Banday,$^{1}$\thanks{banday@MPA-Garching.MPG.DE}
C. Dickinson,$^{2}$\thanks{cdickins@jb.man.ac.uk} R. D. Davies,$^{2}$
R. J. Davis,$^{2}$ and K. M. G\'{o}rski$^{3,4}$\thanks{\emph{Current
address: Jet Propulsion Laboratory, California Institute of
Technology, 4800 Oak Grove Drive, Pasadena CA 91109, USA.}}  \\
$^{1}$Max-Planck Institut f\"{u}r Astrophysik,
Karl-Schwarzschildstrasse 1, 85741 Garching bei
M\"{u}nchen, Germany. \\
$^{2}$Jodrell Bank Observatory, Dept of Physics \& Astronomy,
University of Manchester, Macclesfield, Cheshire SK11 9DL UK. \\
$^{3}$ESO, Karl-Schwarzschildstrasse 2, 85740 Garching bei
M\"{u}nchen, Germany. \\
$^{4}$Warsaw University Observatory, Aleje Ujazdowskie 4, 00-478 Warszawa, Poland.}

\date{Received
**insert**; Accepted **insert**}
       
\pagerange{\pageref{firstpage}--\pageref{lastpage}} \pubyear{}

\maketitle
\label{firstpage}

\begin{abstract}

With the advent of all-sky \ha\ surveys it is possible to determine a
reliable free-free template of the diffuse interstellar medium
(Dickinson, Davies \& Davis 2003) which
can be used in conjunction with the synchrotron and dust templates
to correct CMB observations for diffuse Galactic foregrounds. We have
used the $COBE$-DMR data at 31.5, 53 and 90 GHz and employed
cross-correlation techniques to re-evaluate the foreground
contributions, particularly that due to dust which is known to be partially
correlated with \ha\ (and free-free) emission.

The DMR microwave maps are found to contain, as well as the expected
synchrotron and free-free components, a component tightly correlated
to the 140 \um\ dust maps of DIRBE. At 31.5, 53 and 90 GHz this
emission is $6.3 \pm 0.6$, $2.4 \pm 0.4$ and $2.2 \pm 0.4$ \uk/(MJy
sr$^{-1}$) at 140~\um\ respectively. When corrected for the
contribution from thermal dust following model 7 of Finkbeiner, Davis
\& Schlegel (1999), a strong anomalous dust-correlated emission 
component remains, which is well-fitted
by a frequency spectrum of the form $\nu^{-\beta}$ where 
$\beta\ \sim$ 2.5 in the DMR frequency range; this is the dominant
foreground at 31.5 GHz. 
The result implies the presence of an emission component with a
dust-like morphology but a synchrotron-like
spectrum. We discuss the possible origins of this
component and compare it with the recent {\it WMAP} interpretation.

The better knowledge of the individual foregrounds provided by the
present study enables a larger area of the sky ($|b|>15\degr$) to be used to
re-appraise the CMB quadrupole normalisation, \qrms, and the scalar
perturbations spectral index, $n$. We find
\qrms$=15.2^{+2.8}_{-2.3}$ with a power-law spectral index of
$n=1.2 \pm 0.2$. These values are consistent with previous $COBE$-DMR
analyses and the {\it WMAP} 1-year analysis.

\end{abstract}

\begin{keywords}
cosmic microwave background - cosmological parameters - radio
continuum: general - ISM: dust, extinction   
\end{keywords}


\section{Introduction}
\label{sec:intro}

A major goal of cosmology is to determine the fundamental cosmological
parameters that describe our Universe. Observations of anisotropies in
the cosmic microwave background (CMB) provide unique data in
achieving this goal through a determination of their power
spectrum. An accurate measurement of the CMB power spectrum requires a
good knowledge of the various Galactic foregrounds which have their
own frequency and power spectra. This is particularly important for
all-sky, high-sensitivity CMB experiments such as {\it WMAP} (Bennett
et al. 2003) and
{\it Planck} (Tauber 2000). For example, the CMB power spectrum
expected from observations with the {\it Planck} satellite will be accurate
to $\sim 1$ per cent over the $\ell$-range $100-2000$; this
corresponds to a temperature accuracy of better than $1~\mu$K over the
angular range $2^{\circ}$ to 5 arcmin. The Galactic foregrounds should
therefore be estimated to better than this value, otherwise the CMB
data will be degraded. The $COBE$-DMR data (Bennett et
al. 1996) covering
$\ell=2-30$ at frequencies of 31.5, 53 and 90 GHz are constrained by
knowledge of the foregrounds; this is also the case for the the {\it WMAP}
data, particularly on the largest angular scales discussed here.
The challenge is to make use of the largest sky coverage possible to reduce the sample variance on
such large angular scales ($\gtsim 7^{\circ}$); the area available has
normally been restricted to
Galactic latitudes $|b| \gtsim 20^{\circ}$ or $30^{\circ}$. In this paper we
attempt to improve on available 
knowledge of the foregrounds and apply it to the $COBE$-DMR data
in order to determine whether the fraction of sky available for
cosmological assessment can be extended. The recent {\it WMAP}
analysis of the 1st year observations (Bennett et al. 2003a) uses the
23 GHz channel as a basis for a Galactic mask which eliminates $\sim$21 per
cent of the sky from detailed CMB analysis (Bennett et al. 2003b).

Previous studies of Galactic foregrounds relevant to the CMB at
frequencies less than 100 GHz have used the 408~MHz map of Haslam et
al. (1982) as the synchrotron template and FIR dust emission (IRAS or
DIRBE) as a template for the free-free which was thought to be
correlated with dust and H{\sc i} (Kogut et al. 1996a,b). 
It was widely established that at $10-50$ GHz a dust-correlated foreground was
present (de Oliveira-Costa et al. 1997; Leitch et al. 1997; 
Kogut 1999) although it was at a level which significantly exceeded the
free-free emission expected from the partial \ha\ surveys available at
the time. This dust-correlated anomalous emission has been
ascribed to spinning dust (Draine \& Lazarian 1998a, hereafter DL98, 1998b) 
and appears to fit the expected narrow frequency spectrum 
(de Oliveira-Costa et al. 1998, 1999, 2000, 2002), although Mukherjee et
al. (2001) weaken some of these arguments and claim that
definitive statements about the origin of the dust-correlated
component cannot be made based on data at 10 and 15 GHz alone. Bennett
et al. (2003b) find a dust-correlated component which they interpret
as synchrotron emission with a flatter spectral index. Lagache (2003),
however, argues that this emission is intrinsic to the dust itself and arises
from small, cold grains. This is discussed further in 
section \ref{sec:wmap_comparison}.

A significant  development now available for assessing and quantifying
the Galactic foregrounds is the unambiguous identification of the
free-free component independent of the quasi-correlated dust. The new
\ha\ surveys by the Wisconsin H-Alpha Mapper, WHAM (Reynolds et al. 1998; Haffner 1999), 
and the Southern H-Alpha Sky Survey Atlas, SHASSA (Gaustad et al. 2001), now
make this possible. They can be used to produce an all-sky free-free
template when corrected for \ha\ absorption by dust and when evaluated
at the electron temperature appropriate to the ionized gas in the
local region of the Milky Way (Dickinson et al. 2003).

In the present paper, using cross-correlation techniques, we quantify
the synchrotron and free-free emission, leaving a clearer estimate of
the microwave emission associated with Galactic dust. We use the
$COBE$-DMR data at 31.5, 53 and 90 GHz (Bennett et al. 1996) to
identify an anomalous dust-correlated component which previously
contained an unknown amount of
free-free emission.
Our results are strengthened by including correlations
with a 19.2 GHz survey of the sky (Cottingham 1987; Boughn et al. 1992).

Our present study leads to a frequency spectral index $\beta$ for the
synchrotron emission; although $\beta$ varies with position this
global value is adequate for predicting the actual (small)
contribution at higher radio frequencies ($\nu \gtsim 30$ GHz). We
also investigate the effects of dust absorption and electron
temperature in deriving the free-free template.

Finally, we utilise our improved understanding of the three Galactic templates
to re-examine the determination of the quadrupole normalization
\qrms\ and the scalar fluctuation index \n\ from the
$COBE$-DMR data (G\'orski et al. 1996). We extend the area of sky
available by using data closer to the Galactic plane and examine the
stability of $Q_{\rm rms-PS}$ and $n$ values as a function of the
latitude range included in the analyses.


\section{DMR data and the foreground templates}
\label{sec:data}

\subsection{Synchrotron template}
\label{sec:Synchrotron_template}

The synchrotron emission from the Galaxy originates in relativistic
cosmic ray (CR) electrons spiralling in Galactic magnetic
fields. These electrons will have a a range of spectral energy
distributions depending on their age and the environment of origin
(e.g. supernova explosions or diffuse shocks in the interstellar
medium). The variations in CR electron spectral index will translate
into variations in the frequency spectral index of the synchrotron
continuum emission. Synchrotron emission, with its
temperature spectral index of $\beta=2.4$ to $3.0$ ($T \propto
\nu^{- \beta}$), dominates over thermal bremsstrahlung (free-free)
emission with $\beta=2.15$ at frequencies below $\sim 1$ GHz over
most regions of the sky. Exceptions are the Galactic ridge and the
brighter features of the local Gould Belt system (Dickinson et al. 2003) 

The only all-sky map at sufficiently low frequency to provide a
template of Galactic synchrotron emission is the 408~MHz survey by
Haslam et al. (1982). This has a resolution of 51 arcmin and has a
brightness temperature scale ``absolutely calibrated'' with the 404
MHz $8^{\circ}\!.5 \times 6^{\circ}\!.5$ survey of Pauliny-Toth \&
Shakeshaft (1962).

In the present investigation we will derive (for the sky area
analysed) a mean spectral index between 408~MHz and 19.2 GHz from the
Cottingham (1987) survey and 31.5, 53 and 90 GHz from the $COBE$-DMR
survey. It is important to remember that the spectral index of
Galactic synchrotron emission varies with frequency and position on
the sky. The analysis of low-frequency surveys by Lawson et al. (1987)
indicates that the brighter regions away from the Galactic plane have
typical spectral indices of $\beta=2.55$ and $2.8$ at 100 and 800
MHz respectively. Furthermore, the value of $\beta$ evaluated between any
pair of frequencies separated by a factor of $2-3$ varies typically by
$\Delta \beta= \pm 0.15$ over the sky (Reich \& Reich 1988). Bennett
et al. (2003b) propose a model using {\it WMAP} data which includes a flat
spectrum synchrotron component ($\beta \sim 2.5$) in regions of star formation
near the Galactic plane and a steep spectrum component
($\beta \sim 3.0$) from the Galactic halo.

At frequencies higher than $2-3$ GHz no reliable zero levels can be
measured for large-area surveys at intermediate and high latitudes
where CMB measurements are normally conducted and as a consequence,
large area spectral index maps cannot be made. However, the spectral
index of smaller angular scale ($\sim 5^{\circ}$ and less) structures
can be estimated from cross-correlation analyses of maps at different
frequencies. For example, Hancock et al. (1997) found a spectral index
of $\beta=3.0$ at 10 GHz from bremsstrahlung data at 10,15 and 33
GHz. It is interesting to note that the steepening of the brightness
temperature spectral index by 0.5 at higher frequencies expected from
synchrotron losses would lead to $\beta=3.1$ at higher frequencies;
this is the value at these frequencies expected from the local CR
electron spectrum (Platania et al. 1998).

\subsection{\ha\ free-free template}
\label{sec:ff_template}

For the free-free template, we use the all-sky \ha\ template described
in Dickinson et al. (2003). The all-sky map is a composite of \ha\ data
from the WHAM survey of the northern sky
(Reynolds et al. 1998; Haffner 1999) and SHASSA survey in the southern
sky (Gaustad et al. 2001). The SHASSA data are first smoothed to the WHAM resolution of $1^{\circ}$ and are 
preferred at declinations $\delta < -15^{\circ}$. It should be noted that the
all-sky template is a first attempt at producing a high-sensitivity
all-sky \ha\ map. SHASSA data have significantly lower sensitivity than data from the
WHAM survey; the noise levels are estimated as $0.5$ and $0.05$ Rayleigh ($R$) respectively, where $1~R \equiv 10^{6}/4\pi $
photons s$^{-1}$ cm$^{-2}$~sr$^{-1} \equiv 2.41 \times 10^{7}$ erg
s$^{-1}$ cm$^{-2}$~sr$^{-1} \equiv 2.0~ {\rm cm}^{-6} {\rm pc}$ for
$T_{e}=7000$~K gas. Baseline uncertainties exist in both data-sets,
particularly SHASSA data which results in loss of information on angular
scales $\gtsim 10^{\circ}$ for data at $\delta \ltsim -30^{\circ}$ due
to the $13\degr$ field-of-view of each pointing. However, the SHASSA background levels are fitted to
WHAM data where WHAM data exist ($\delta >-30\degr$) with an assumed cosecant
law for declinations further south. The different observing strategies result
in different beam shapes which have not been accounted
for. Finally, large \ha\ signals (emission and absorption) from bright stars have been filtered
using different techniques; the WHAM team use data from nearby
unaffected pixels, while the SHASSA team use a median filter to reduce
star residuals. However, all these possible systematics are
significantly reduced when the data, at a resolution of $\sim 1^{\circ}$,
are convolved with the DMR beam at $\sim 7^{\circ}$ resolution. 
Indeed, we have
explicitly verified this by comparing the power spectra of the high-latitude \ha\
emission after varying the latitude at which the surveys are
matched and adjusting baselines. Recently, Finkbeiner (2003)
has produced an all-sky \ha\ map at 6 arcmin resolution which includes
higher resolution data from the Virginia-Tech Sky Survey (Dennison et
al. 1998). The differences in the power spectrum of our \ha\ map with
that given by Finkbeiner (2003) are less than 10 per cent for $\ell <
15$ and less than 30 per cent everywhere.

The optical \ha\ line ($\lambda = 656.3$ nm) is well known to be a good
tracer of the free-free continuum emission at radio wavelengths since they
are both emitted by the same ionized gas. Since they are both
proportional to the Emission Measure ($EM \equiv \int n_{e}^{2} dl$)
there is a well-defined relationship between the \ha\ line intensity
$I_{\rm H\alpha}$ and the radio continuum brightness temperature
$T_{b}^{ff}$ (Osterbrock 1989; Valls-Gabaud 1998; Dickinson et al. 2003). There is a strong dependence on frequency since
free-free emission scales as $\nu^{2.15}$ at GHz frequencies. There
is also a modest dependence on the electron temperature $T_{e}$
($\propto T_{e}^{0.7}$). In the Galaxy, $T_{e}$ can range between 3000~K and
15000~K (Shaver et al. 1983); the majority of
H{\sc ii} regions have $T_{e}$ between 5000 and 10000~K (Reynolds 1985). 
Shaver et al. (1983) show that the mean electron temperature of
H{\sc ii} regions at the galactocentric radius of the Sun ($R_{0}=8.5$
kpc) is 7200~K with an r.m.s. spread of 1200~K. In the region of the
Galaxy covered in the present study where $|b|>15\degr$ the ionized
hydrogen will be typically within $\sim 1$ kpc of the Sun and accordingly we
adopt a value of $T_{e}=7000$~K. With this value of $T_{e}$, the conversions
from \ha\ to free-free continuum for frequencies 31.5, 53 and 90 GHz
are 5.25, 1.71 and 0.54 $\mu$K~$R^{-1}$ respectively. In section
\ref{sec:results_foregrounds} we will use this prior information to
allow a fixed free-free component to be subtracted giving a clearer
separation of the foregrounds. We will take a variation of $\pm 2000$
K from the local value of 7000~K as in indication of the upper and
lower bounds allowed. Other caveats
become important at the few per cent level: i) the helium abundance is
usually taken to be 8 per cent with a spread of a few per cent ii) the
emitting medium is assumed to be optically thick (case B) to the Lyman
continuum photons.

A major uncertainty when using the \ha\ template is the absorption of
\ha\ by foreground dust. The absorption can be estimated using the $100~\mu$m maps
given by Schlegel, Finkbeiner \& Davis (1998, hereafter SFD). The absorption at \ha\
$A({\rm H \alpha})$ in magnitudes using their temperature-corrected
column density map ($D^{T}$) in units of MJy~sr$^{-1}$ is then $A({\rm
H \alpha}) = (0.0462 \pm 0.0035) D^{T} f_{d}$ where $f_{d}$ is the
fraction of dust {\it in front} of the \ha\ actually causing
absorption in the line of sight. The corrected \ha\ intensity is then $I_{{\rm
H}\alpha}^{\rm corr} = I_{{\rm H}\alpha} \times 10^{D^{T} \times 0.4 \times
f_{d} \times 0.0462}$. Although there are uncertainties in this relationship due
to different reddening values expected from different dust
populations, the largest uncertainty is in $f_{d}$, the fraction of dust lying in
front of the \ha-emitting region. Dickinson et al. (2003) show that
for a range of longitudes ($l=30^{\circ}-60^{\circ}$), and for
$|b|=5^{\circ}-15^{\circ}$, $f_{d}\approx 0.3$, while for local high
latitude regions such as Orion and the Gum Nebula, there is little
or no absorption by dust ($f_{d} \sim 0$). For much of the high
latitude sky, $D^{T}$ is $\ltsim 5$ MJy~sr$^{-1}$ and therefore the
absorption is likely to be small ($\ltsim 0.2$ mag). At lower Galactic
latitudes $|b| \ltsim 10^{\circ}$, the dust absorption strongly increases to 1 mag and above. For the majority of the Galactic plane
($|b|<5^{\circ}$), the absorption is too great to make any reliable
estimate of the true \ha\ intensity.  

Finally, the scattering of \ha\ light from dust at high Galactic
latitudes could, in principle, cause a degeneracy in the analysis
since the scattered \ha\ light would be directly correlated with the
dust. However, Wood \&
Reynolds (1999) predict that this accounts for less than 20 per cent of high
latitude \ha\ emission. Since the anomalous component is larger than
the free-free component at frequencies higher than $\sim 20$ GHz (see
sections \ref{sec:results_foregrounds} and
\ref{sec:results_anomalous}), this systematic will not
significantly affect our results within the errors.     

\subsection{Dust templates}
\label{sec:dust_templates}

As templates for the thermal and anomalous dust-correlated emission, we utilise
the $COBE$-DIRBE data repixelised into the HEALPix
pixelisation scheme. This was achieved using the publicly
available Calibrated Individual Observations (CIO) files together
with the DIRBE Sky and Zodi Atlas (DSZA, both products
being described in Hauser et al. 1998), allowing the
construction of DIRBE full sky maps corrected for the
zodiacal emission according to the model of Kelsall et al. (1998).
In particular, we use the DIRBE 140~\um\ data at $0^{\circ}\!.7$ resolution: although the 100~\um\ 
data has the best sensitivity, there remain more noticeable artifacts
present in the sky map after zodiacal correction. While
this is not likely to present a problem after the data has
been convolved with the DMR beam profile ($\sim 7^{\circ}$ FWHM), our choice allows 
maximal consistency with previous analyses by the DMR team 
(Banday et al. 1996; G\'orski et al. 1996; Kogut et al. 1996a, 1996b).
Furthermore, the SFD version of the 100~\um\ sky map has
been used to correct for dust absorption in the \ha\ template,
thus our choice allows some additional 
independence in the dust correlation results. 

DL98 have suggested that the anomalous dust-correlated signal
which we re-investigate in this work should be significantly correlated
with 12 \um\ emission, so that the corresponding 
DIRBE sky map may prove a better tracer for this foreground
component. However, the 12 \um\ map contains residual zodiacal artifacts
in the region of the ecliptic plane which remain
significant even after smoothing to DMR resolution. The impact 
of this could be minimised by excluding a region close to the
ecliptic plane, but we prefer to make use of the greatest sky
coverage possible. Further, de Oliveira-Costa et al. (2002)
have tested this hypothesis, but it is clear from their
fig. 4 that even after considerable processing of the
12 \um\ DIRBE data for the removal of both zodiacal emission and 
point sources, the longer wavelength data at 100, 140 and 240 \um\
remain equally viable for correlation studies.

Finally, in order to attempt to isolate the anomalous dust-correlated
component signal, it is necessary to identify and subtract the thermal
dust emission. We have employed the predictions of
Finkbeiner, Davis \& Schlegel (1999, hereafter FDS)
for the thermal dust emission at DMR wavelengths. FDS consider
8 models based on the 100 and 240 \um\ maps from SFD, and test
them against the $COBE$-FIRAS data. The inconsistency of
extrapolations of the SFD maps using single component 
power-law emissivity functions with the FIRAS data over 
the range 200 - 2100 GHz motivated the development of
generalised two-component models. We have considered FDS models
7 and 8 as plausible tracers of the thermal dust component 
emission. Masi et al. (2001) have demonstrated that 
model 8 extrapolated to 410 GHz is well correlated with the BOOMERanG 
data at that frequency, 
and a recent analysis of the MAXIMA data (Jaffe et al. 2003)
also see consistency with these extrapolated dust models.
It should be remembered that the FDS models are formally 
correct only on scales comparable to or larger than the FIRAS
beam, 
hence the FDS predictions of model 7/8 should be well-matched 
to our purposes.

\subsection{$COBE$-DMR data}
\label{sec:dmr_data}

In this analysis, we utilise the $COBE$-DMR 4-year sky maps (Bennett
et al. 1996) at 31.5, 53 and 90 GHz in Galactic coordinates 
and the HEALPix\footnote{\emph{http://www.eso.org/science/healpix/}}
pixelisation scheme with a resolution parameter of $N_{\mathrm{side}} = 32$,
corresponding to 12\,288 pixels on the sky. 
Inverse-noise-variance weighted combinations of the A and B channels 
are formed at each of the three DMR frequencies, the weights being 
(0.618, 0.382), (0.579, 0.421) and (0.382, 0.618) at 31.5, 53 and 
90 GHz respectively. The dominant feature visible in the sky maps
(given that the dipole anisotropy of amplitude $\sim$ 3 mK 
is largely removed in the pre-map-making data processing)
is the enhanced emission along the Galactic plane. This emission
is indeed a combination of the synchrotron, free-free and dust emission
that we are interested in, however, the distribution and spectral
behaviour at the lowest latitudes ($|b| < 10\degr$) is sufficiently complex that
we do not attempt to fit or model it, considering only the
mid- and high-latitude Galactic emission. Contrary to previous
analyses, we attempt to fit our templates to somewhat lower 
latitudes, the results of which are discussed 
in section~\ref{sec:results_foregrounds}.

\subsection{19.2 GHz data}
\label{sec:19ghz_data}

In addition to the $COBE$-DMR data, we also consider correlations
of the foreground templates with a 19.2 GHz survey of the sky
(Cottingham 1987; Boughn et al. 1992).
Such an investigation is clearly of interest here since 
19.2 GHz is close to the frequency at which the DL98 models for spinning
dust show a maximum in spectral behaviour. Even if the anomalous dust-correlated component is
not the putative rotational dust emission, the coupling amplitude of the dust 
template at 19.2 GHz should enable the spectrum of this foreground to be
established, in particular since the frequency of observation is well-removed
from that where a significant thermal dust contribution is present. The
data are available in the $COBE$ QuadCube format (White \& Stemwedel, 1992)
with 24,576 pixels on the full sky of size $1^{\circ}.3\, \times\,
1^{\circ}.3$, and contains observations of angular resolution 3\degr\ FWHM. A small region close
to the South Ecliptic pole was not observed. The typical noise level
is quoted as 1.5 mK per resolution element, but there is both
variation in the number of observations per pixel and pixel-pixel
covariance. In contrast to the analysis of de Oliveira-Costa et al. (1998)
we take account of the former, but the information is not available for
the latter. Some attempt to provide additional constraints on the 
noise uncertainty is performed by scaling the noise level by a factor
in the range 0.9 - 1.1, and including this parameter in
our later statistical analysis.


\section{Analysis method}
\label{sec:method}

Each of the three DMR maps, the 19.2 GHz data,  and the foreground sky map
templates (after convolution to the correct angular resolution) 
are decomposed into harmonic coefficients using a basis of 
orthonormal functions explicitly computed 
on the incompletely sampled sky (G\'orski 1994), a procedure which is
directly analogous to the expansion of a full-sky map
in terms of spherical harmonic coefficients.
This technique has the advantage of allowing exact exclusion 
of the monopole and dipole components 
from the analysis (which are unimportant for the investigation
of cosmological anisotropy), and makes full use of all of the available 
spatial (phase) information.

We have adopted several definitions for the area of sky coverage, in
particular the standard Galactic cut 
defined as the region with Galactic latitude, $|b|$,
above 20$\degr$ with additional regions at Ophiuchus (up to $b \approx
30\degr$ at $l \approx 0\degr$) and Orion (down to $b \approx
-35\degr$ at $l \approx 180\degr$) based
on the DIRBE 140~\um\ map
(Banday et al. 1997, but recomputed explicitly for the
HEALPix scheme), together with more aggressive
and conservative cuts of $|b|$ greater than 15$\degr$ and 30$\degr$
respectively. For each sky coverage a new basis of orthonormal 
functions is computed. This procedure must also be performed
separately for the 19.2 GHz data due to its different pixelisation
scheme and pixel resolution (and indeed for the foreground
templates matched to the same pixelisation and angular 
resolution).

The measured map coefficients in the new orthogonal basis at 
a given frequency, \cnu, can then 
be written in vector form as
${\mathbf c}_{\nu}\ = {\mathbf c}_{\mathrm CMB}\ +\ 
{\mathbf c}_{\mathrm N}\ 
+\ \alpha_X {\mathbf c}_{X}$ where ${\mathbf c}_{\mathrm CMB}$, 
${\mathbf c}_{\mathrm N}$ and ${\mathbf c}_{X}$ 
are the coefficients describing the true CMB distribution,
the noise and one foreground template map respectively. $\alpha_X$ is now a coupling constant
(with units \uk\ $X^{-1}$) to be determined by minimizing 

\begin{equation}
\chi^{2}\ =\ ({\mathbf c}_{\nu}\ -\ \alpha_X {\mathbf c}_{X})^{T}\, 
{\mathbf\mathrm \tilde{M}}^{-1}\,
             ({\mathbf c}_{\nu}\ -\ \alpha_X {\mathbf c}_{X})
\label{eqn:chi2}
\end{equation}
where {\bf \~{M}} is the covariance matrix 
describing the correlation between different Fourier modes 
on the cut-sky and is dependent on
assumed CMB model parameters and the instrument noise. 
Full details of its computation can be found in
G\'orski (1994) and G\'orski et al. (1996).

By minimising the $\chi^2$ with respect to $\alpha_X$ we find that 
the coupling constant $\alpha_X$ has an exact solution
\begin{equation} 
\alpha_X\ =\ {\mathbf c}_{X}\, {\mathbf\mathrm \tilde{M}}^{-1}\, 
{\mathbf c}_{\nu}/
           {\mathbf c}_{X}\, {\mathbf\mathrm \tilde{M}}^{-1}\, {\mathbf c}_{X}
\end{equation}
and a Gaussian error given by 
\begin{equation}
\sigma^{2}(\alpha_X)$ =\
  $({\mathbf c}_{X}\, {\mathbf\mathrm \tilde{M}}^{-1}\, {\mathbf c}_{X})^{-1}
\end{equation}

The contribution of component $X$ to the observed sky is the map $\alpha_X T_X$
(where $T_X$ is in the natural units of the foreground template map
so that this scaled quantity is in \uk).

In this analysis, however, we treat multiple foreground templates
and all three DMR channels simultaneously. 
This forces the CMB distribution (in thermodynamic units)
to be invariant between the three frequencies, 
and suppresses the sensitivity of the method to 
noise features in the individual DMR maps which are clearly not
common to all three frequencies,
thus allowing an improved determination of the extent of chance
cosmic or noise alignments with the template maps.
We have tested the impact of such chance alignments to the method by repeating
the fits after rotation of the foreground templates. We only find significant correlations when the templates are correctly
aligned. Moreover, the r.m.s values of the coupling coefficients derived from
the rotated sky maps are consistent with the errors determined 
directly in the analysis.
The full details of this
method are contained in G\'orski et al. (1996). 
The 19.2 GHz data are treated in a similar
fashion, but we do not extend the method to include it in a simultaneous
fit with the DMR data to the foreground templates because of complications 
introduced by its different angular resolution and noise properties.

Having solved for the values of the coupling coefficients and
their Gaussian errors, we are then in a position to compute
the likelihood of the foreground corrected data as a function
of a given set of input cosmological parameters as contained
in the matrix $\mathbf\mathrm \tilde{M}$ of equation~\ref{eqn:chi2}.
Indeed, this also allows us to assess the sensitivity of the 
coupling constant fits to the assumed cosmological power spectrum.
We consider a full grid of power law models specified by the
rms quadrupole normalisation amplitude \qrms\ and a spectral
index \n\, as in G\'orski et
al. (1996). The cosmological implications of the 
foreground corrections as reappraised in this work are discussed
in section~\ref{sec:results_CMB}.

In order to quantify the uncertainty in the \ha\ template resulting
from dust absorption, we generate corrected \ha\ templates  
for three values of $f_{d}$ -- 0.0, 0.5 and 1.0 -- which are
then fitted successively to the data over the full 
{\qrms, \n} grid of cosmological models, and also test
templates corrected assuming values of $f_{d}$ running 
from 0 to 1 in steps of 0.01 for a scale-invariant spectrum
with \qrms~$\sim$18~\uk. Since the correction is non-linear in
$f_{d}$ (see section~\ref{sec:ff_template}), modifications to the template are made
\emph{before} convolution with the appropriate
beam response, at the 1\degr\ effective angular resolution of the
\ha\ template (see section~\ref{sec:results_dmr_ff} for full discussion). 

We also analyse the DMR and 19.2 GHz data after correction for 
free-free emission using the \ha\ templates scaled by the factors
given in Table~4 of Dickinson et al. (2003). This enables us to test
the impact of cross-talk between the templates, which is likely to be
present at some level due to the large-scale cosecant variation with 
latitude of the Galactic emission present in
each template (see section~\ref{sec:results_19ghz}). 

Finally, in an attempt to unambiguously identify the \emph{anomalous}
dust-correlated emission, we use the FDS models of \emph{thermal} dust emission
and subtract them from the observed data before correlation studies
are performed (see section~\ref{sec:results_anomalous}).


\section{Results - foregrounds}
\label{sec:results_foregrounds}

\begin{figure*}
\vbox to305mm{\vfil \includegraphics[angle=90]{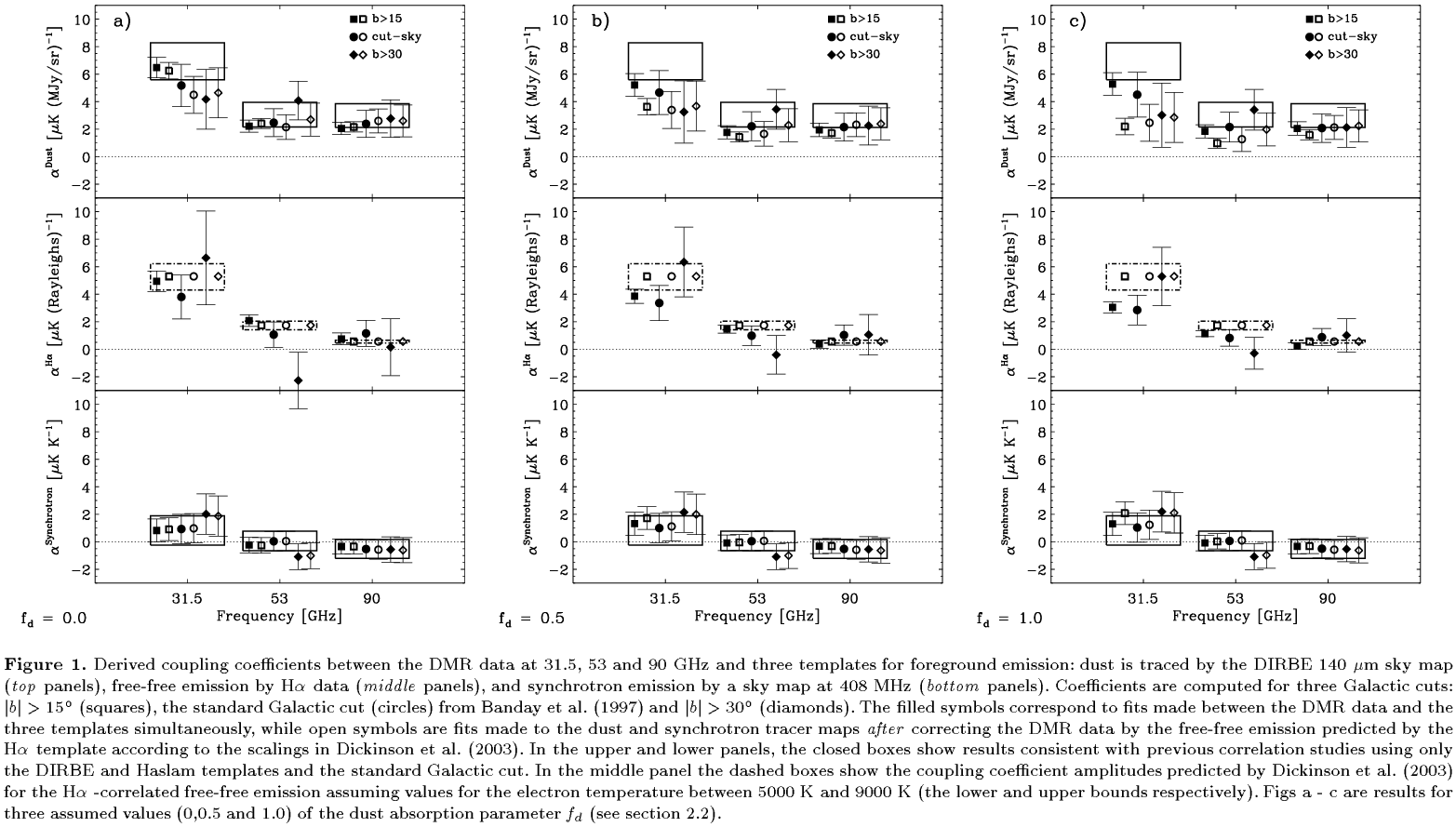}  
\vfil}
\caption{}
 \label{fig:param_fits_fd} 
\end{figure*}

The results of our correlation study for the three DMR frequencies are
summarised in Figs \ref{fig:param_fits_fd}(a)-(c). This summary gives
information for six parameters as follows:-

1. Frequency. Data are grouped in each of the three figures (a)-(c)
   as three bands at each of the DMR frequencies 31.5, 53 and 90 GHz.

2. Foreground components. The three panels in each figure give the
   results for dust, free-free and synchrotron plotted from top to
   bottom.

3. Galactic cut. The results are shown for three Galactic cuts, namely
   $|b|>15\degr$ (squares), the standard Galactic cut of Banday et
   al. (1997) (circles) and $|b|>30\degr$ (diamonds).

4. \ha\ extinction. The effect of the three levels of \ha\ extinction,
   $f_{d}=0,0.5$ and $1.0$, are shown in Figs (a), (b) and (c)
   respectively.

5. Forced/unforced fits to the \ha\ data. Open symbols are fits that
   are made to the dust and synchrotron templates \emph{after}
   correcting the DMR data for the free-free contribution predicted by the \ha\
   template; filled symbols correspond to fits made to the three templates
   simultaneously in which no constraints have been placed on the \ha\
   template correlation.

6. Electron temperature. The free-free predictions from the \ha\
   template depend upon the electron temperature; the dashed boxes in
   the middle panels show the range of predicted free-free emission
   for temperatures 5000 to 9000~K (open symbols are for 7000~K).

   The results obtained
   from an analysis using only the DIRBE and synchrotron
   templates, consistent with previous work by the $COBE$-DMR team
   (Kogut et al. 1996b), are shown with their errors by full-line boxes in the
   upper and lower panels of each figure\footnote{Note
   that the boxes in the top and bottom panels of
   Fig.~\ref{fig:param_fits_fd} show slightly different results to the previous
   analysis of Kogut et al. (1996b). However, this is not surprising
   since they used an earlier release of
   the DIRBE 140~\um\ map and the data were analysed in the QuadCube
   format.}. This allows a direct assessment of the impact of the
   inclusion of a free-free template on the analysis by comparison
   with results in which such a template has not been utilised.

The results presented here have been derived assuming a
scale-invariant spectrum (\n\ = 1) for the CMB contribution with an r.m.s
quadrupole normalisation of \qrms\ = 18 \uk. However, as part of our
studies of the impact of the foreground correction for cosmological
interpretations of the DMR data, we have determined the coupling
parameters over a grid in \qrms\ and \n. 
We compute an average of the $\alpha$ coefficients 
weighted by the likelihood function over the grid and
find that the marginalised coupling
coefficients are essentially identical to those shown here, and that
no conclusions derived from later analysis are compromised by adopting
these values as canonical. As described above, we have adopted three
Galactic cuts, and also computed correlations in which the coupling to
the \ha\ template has been left unconstrained, or where we have fixed
and subtracted the predicted free-free contribution according to the
spectral behaviour predicted in Dickinson et al. (2003), determined
for an electron temperature of 7000~K, the mean value in the
solar neighbourhood (see section~\ref{sec:ff_template}).

The following sections provide a detailed discussion of this data
set. 

\subsection{DMR and Dust}
\label{sec:results_dmr_dust}

\begin{table*}
  \begin{center}
\caption[]{The predicted foreground r.m.s signal in the DMR sky maps
  at $\sim$ 7\degr resolution
  due to dust, free-free and synchrotron emission, determined by
  scaling the corresponding emission templates by the best-fit
  coupling coefficients. Quoted errors are 68\% confidence level, all
  values are in \uk\ and have been converted to thermodynamic
temperature relative to a fixed CMB temperature,  $T_{\rm CMB}=2.73$~K . We do not distinguish between the thermal and
  anomalous dust-correlated contributions. Note the lack of sensitivity to \fd\
  for the free-free contribution.}
  \label{tab:rms_signal}  
  \begin{tabular}{llccccccccc} \hline
    Component & Coverage & \multicolumn{9}{c}{DMR Map (GHz)} \\ \cline{3-11}
              &          & \multicolumn{3}{c}{31.5} & \multicolumn{3}{c}{53} & \multicolumn{3}{c}{90} \\ \cline{3-11}
               & & $f_d\, = $ 0.0 & 0.5 & 1.0 & $f_d\, = $0.0 & 0.5 & 1.0 & $f_d\, = $0.0 & 0.5 & 1.0 \\ \hline\hline
             & $b>15\degr$ &  51.8$\pm $ 6.0 &  41.6$\pm $ 6.6 &  42.2$\pm $ 6.6 &  18.6$\pm $ 3.7 &  14.8$\pm $ 4.1 &  15.4$\pm $ 4.0 &  19.6$\pm $ 4.3 &  18.6$\pm $ 4.8 &  19.6$\pm $ 4.7 \\
        Dust &    standard &  19.9$\pm $ 5.9 &  17.9$\pm $ 6.1 &  17.3$\pm $ 6.3 &  10.0$\pm $ 4.1 &   8.9$\pm $ 4.3 &   8.7$\pm $ 4.3 &  11.0$\pm $ 4.5 &   9.9$\pm $ 4.7 &   9.6$\pm $ 4.8 \\
             & $b>30\degr$ &  13.6$\pm $ 7.1 &  10.6$\pm $ 7.4 &
9.9$\pm $ 7.6 &  14.0$\pm $ 4.8 &  11.8$\pm $ 4.9 &  11.7$\pm $ 5.1 &
10.8$\pm $ 5.3 &   8.8$\pm $ 5.5 &   8.3$\pm $ 5.7 \\ \hline
             & $b>15\degr$ &  29.1$\pm $ 4.4 &  35.8$\pm $ 4.8 &  35.4$\pm $ 4.8 &  12.9$\pm $ 2.6 &  14.2$\pm $ 2.8 &  13.8$\pm $ 2.7 &   5.3$\pm $ 3.0 &   4.1$\pm $ 3.3 &   3.4$\pm $ 3.3 \\
   Free-free &    standard &  10.0$\pm $ 4.2 &  11.2$\pm $ 4.2 &  11.5$\pm $ 4.3 &   2.9$\pm $ 2.5 &   3.5$\pm $ 2.5 &   3.4$\pm $ 2.5 &   3.6$\pm $ 2.9 &   4.1$\pm $ 2.9 &   4.2$\pm $ 3.0 \\
             & $b>30\degr$ &  10.8$\pm $ 5.6 &  13.6$\pm $ 5.4 &
14.1$\pm $ 5.7 &  -3.9$\pm $ 3.5 &  -0.9$\pm $ 3.1 &  -0.8$\pm $ 3.2 &
0.3$\pm $ 4.0 &   2.7$\pm $ 3.8 &   3.2$\pm $ 3.9 \\ \hline
             & $b>15\degr$ &   5.3$\pm $ 5.4 &   8.5$\pm $ 5.4 &   8.5$\pm $ 5.5 &  -1.6$\pm $ 3.8 &  -0.5$\pm $ 3.8 &  -0.6$\pm $ 3.8 &  -2.7$\pm $ 4.2 &  -2.5$\pm $ 4.2 &  -2.6$\pm $ 4.2 \\
 Synchrotron &    standard &   4.7$\pm $ 5.4 &   5.0$\pm $ 5.4 &   5.2$\pm $ 5.4 &   0.2$\pm $ 3.7 &   0.2$\pm $ 3.7 &   0.3$\pm $ 3.7 &  -3.2$\pm $ 4.1 &  -3.1$\pm $ 4.1 &  -3.0$\pm $ 4.1 \\
             & $b>30\degr$ &   8.0$\pm $ 5.9 &   8.6$\pm $ 5.9 &
8.7$\pm $ 5.9 &  -4.5$\pm $ 3.9 &  -4.5$\pm $ 3.9 &  -4.5$\pm $ 3.9 &
-2.7$\pm $ 4.4 &  -2.6$\pm $ 4.4 &  -2.5$\pm $ 4.4 \\ \hline
  \end{tabular}
  \end{center}
  \label{tab:}
\end{table*}

In order to determine accurate values of the coupling amplitudes
between the DMR data and the three foreground templates it is
important that the templates are uncorrelated or largely
uncorrelated, or that we know the precise emission expected from one
of the partially correlated templates. The extent of the partial
correlation between the dust and the free-free (\ha) emission can be
elucidated in the present study by first removing the free-free emission
derived from the \ha\ template (section~\ref{sec:ff_template}) before
deriving the coupling coefficient for DIRBE dust. The computation was
also made without this \ha\ constraint; the results were not
significantly different. It is also necessary to correct the \ha\
template for absorption by intervening dust by an absorption factor
\fd; the analysis of Dickinson et al. (2003) indicates that this is
between 0 and 0.3. At high Galactic latitudes this has little effect
since the total absorption is small ($A($\ha$ \ltsim 0.1$). The plots in Fig.~\ref{fig:param_fits_fd} show the
effect of assuming different values of \fd. In fact, the stability of
the dust couplings derived when the free-free coupling is
unconstrained implies that there is minimal cross-talk between the
components. The results when the free-free coupling is fixed are more
strongly influenced by the relation between $T_{e}$ and $f_{d}$ as
discussed in more detail in section \ref{sec:results_dmr_ff}.

The estimates of the coupling amplitudes of DMR data against the DIRBE
140~\um\ template at the the three frequencies of DMR are given in the
top panels of Fig.~\ref{fig:param_fits_fd}a. The best estimates are
for \fd=0 (see discussion in section~\ref{sec:results_dmr_ff}), for the largest sky area
covered ($|b| \gtsim 15\degr$) and for a fixed \ha\ contribution. The values are $6.3
\pm 0.6$, $2.4 \pm 0.4$, $2.2 \pm 0.4$ \uk/(MJy~sr$^{-1}$) at 31.5, 53
and 90 GHz respectively. We compare these values with those of Kogut
et al. (1996b) using only the DIRBE and the Haslam templates of $6.4
\pm 1.5$, $2.7 \pm 1.1$ and $2.8 \pm 1.0$ \uk/(MJy~sr$^{-1}$). There is a small reduction in
coupling amplitude when the free-free (\ha) template is taken into
account. Significantly, there is a factor of $2-3$ reduction in the
r.m.s. error when the free-free emission is included in the
analysis. This reduction is partly due to the larger area of sky covered but is mainly due to the substantial scatter in the
correlation between dust and \ha\ emission (Dickinson et al. 2003) and
the fact that they are individually cross-correlated with the DMR
data. 

We will now consider the r.m.s. brightness temperature contribution
expected for each foreground component as calculated from the
correlation coefficients in Fig.~\ref{fig:param_fits_fd}. The entries
in Table \ref{tab:rms_signal} are the product of the correlation
coefficient and the r.m.s. level on each of the template maps 
(after subtraction of the best-fitting monopole and dipole)
for the relevant range of Galactic latitudes 
and for each of the three DMR
frequencies. In Table \ref{tab:rms_signal} we have converted the
conventional (Rayleigh-Jeans) brightness temperature units into
thermodynamic temperature units (relative to a fixed CMB temperature, $T_{CMB}$) by
multiplying by the Planck correction factor
$(e^{x}-1)^{2}/(x^{2}e^{x})$ where $x=h \nu/kT_{\rm CMB}$; the factors
are $\sim$ 1.03, 1.07 and 1.23 at 31.5,53 and 90 GHz respectively. At the preferred value of $f_{d}$=0, the
rms dust emission at 31.5 GHz for the $|b|>15\degr$ cut is $\sim 4$
times that for $|b|>30\degr$ reflecting the strong increase towards
the Galactic plane. The table shows that the dust emission is $\sim
1.5$ times stronger than the free-free, the next brightest
component. We note that the r.m.s. error in the dust foreground
estimate is similar ($\sim 6~\mu$K) for each Galactic cut; this
suggests that the precision with which the CMB can be corrected for
dust emission is not degraded on extending the field of view to $|b|
\sim 15\degr$.

\subsection{DMR and Synchrotron emission}
\label{sec:results_dmr_synch}

Although synchrotron emission is weak at DMR frequencies, some
significant conclusions can be drawn from the cross-correlation
between the DMR data and the 408~MHz synchrotron template. The
most compelling data are shown in the bottom panel of
Fig.~\ref{fig:param_fits_fd}(a) which is for $f_{d}=0.0$, and for the
cross-correlation obtained with the free-free emission predicted from
the \ha\ template subtracted, as shown by the open squares. The
coupling constants $\alpha^{Synchrotron}$ for the largest sky
coverage ($|b|>15\degr$), are $0.91 \pm
0.83$, $-0.26 \pm 0.56$ and $-0.35 \pm 0.54$ \uk\, K$^{-1}$ at 31.5, 53
and 90 GHz respectively. 
The coupling constant can be converted to limits on the 
brightness temperature spectral index $\beta$ between 0.408 GHz and the observing
frequency, $\nu$, using $\alpha^{Synchrotron} =
(\frac{\nu}{0.408})^{-\beta}$.

\begin{table}
  \begin{center}
  \caption[Synchrotron spectral index]{The derived synchrotron
           spectral index between 408~MHz and the quoted frequency,
           assuming the power-law relation
           $\alpha^{\mathrm Synchrotron}= \left(\frac{\nu}{0.408}\right)^{-\beta}$.
           Quoted errors are 68\% confidence level, upper limits are
           at the 95\% confidence level. The numbers are determined
           from the coupling amplitudes resulting from an analysis adopting a fixed
           free-free contribution.}
  \label{tab:synch_index}   
  \begin{tabular}{cccc} \hline
  Frequency (GHz) & \multicolumn{3}{c}{Sky Coverage} \\ \cline{2-4}
                  & $|b| > 15\degr$   & Standard        & $|b| > 30\degr$   \\ \hline\hline
  19.2            & 3.08$^{+0.10}_{-0.07}$ & 3.02$^{+0.10}_{-0.07}$ & $>$3.08 \\
  31.5            & $>$3.07         & $>$3.01         & $>$2.93         \\
  53.0            & $>$2.83         & $>$2.78         & $>$2.72         \\
  90.0            & $>$2.58         & $>$2.54         & $>$2.49         \\ \hline  
  \end{tabular}
  \end{center}
\end{table}

Table \ref{tab:synch_index} gives the values of $\beta$ for the three
sky cuts used in the present study. The most significant results are
for the $|b|>15\degr$ cut which includes the brightest backgrounds;
$\sim 95\%$ ($2\sigma$) upper limits are given for $\beta$. The 19.2 GHz
result will be discussed in section
\ref{sec:results_19ghz}.

The spectral index calculated between 0.408 GHz and 20 to 30 GHz where
synchrotron signals can be detected is $\beta=3.1$. $\beta$ is known
to increase with frequency between $\sim 0.1$ and 1 GHz (see section
\ref{sec:Synchrotron_template}) with typical values of 2.7 at 1
GHz. Clearly $\beta$ steepens further by $20-30$ GHz; the ``local''
value of the index will be greater than 3.1 at DMR frequencies if the
synchrotron-loss turnover ($\Delta \beta=0.5$) occurs at frequencies
higher than a few GHz. Spectral indices of $\sim 3$ were considered
by Banday \& Wolfendale (1991). Direct measurements at frequencies of
$5,10$ and 15 GHz in the Tenerife and Jodrell Bank experiments (Jones
et al. 2001, Hancock et al. 1997) indicate spectral indices of $\sim 3$ between 0.408 GHz and these
frequencies. We note that the values derived in Table
\ref{tab:synch_index} are the average spectral index over a
substantial part of the sky; there is presumably a scatter of
spectral index at high frequencies corresponding to that found at low
frequencies. 

Table \ref{tab:rms_signal} shows that the r.m.s. synchrotron emission
in the DMR data is essentially the same for each of the three latitude cuts
in contrast to the dust and free-free emission which increases
significantly towards the Galactic plane. This is because in the 408
MHz synchrotron template map, the North Polar Spur (NPS) dominates the
emission in each latitude cut. The NPS is the major contribution to
the synchrotron r.m.s. temperature at these frequencies. This will be
discussed further in section~\ref{sec:results_19ghz}.

\subsection{DMR and Free-free}
\label{sec:results_dmr_ff}

From Fig.~\ref{fig:param_fits_fd}, the \ha\ template appears to be well correlated against the DMR data at
all frequencies for the 15\degr Galactic cut, and at better than the
2$\sigma$ level at 31.5 GHz for each of the cuts. The derived
correlations with the \ha\ template are completely consistent with that expected for
free-free emission due to thermal electrons with temperatures in the
range $5000 - 9000$~K. For the larger values of \fd, the picture is less
consistent, and it appears that some of the \ha\ correlation is at the
expense of the dust, particularly for the case of fixed
scaling. For a value of \fd\ of order unity, the derived scaling for a
15\degr cut is inconsistent with this temperature range.

\setcounter{figure}{1}
\begin{figure}
  \begin{center}
    \includegraphics[width=0.5\textwidth]{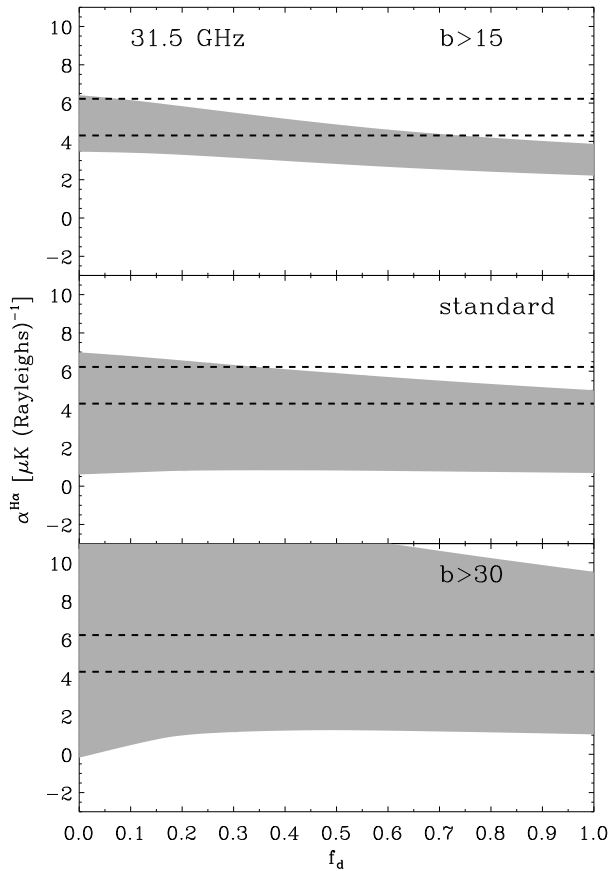}
  \end{center}
  \caption{The coupling coefficient determined between the DMR data at
    31.5 GHz and the \ha\ tracer of free-free emission as a
    function of the dust absorption parameter \fd. The grey bands
    correspond to the 95\% confidence ranges for the derived coupling
    coefficient amplitudes. The dotted lines are the 
    predictions of Dickinson et al. (2003) for the
    \ha-correlated free-free emission assuming values for the electron
    temperature between 5000~K and 9000~K (the lower and upper bounds
    respectively). There is weak evidence that the data prefer low
    values for \fd, moreover if we consider our \lq preferred' value
    for the electron temperature of 7000~K it appears that 
    \fd\ is constrained to be less than $\sim 0.35$.}
  \label{fig:full_fits_fd}
\end{figure}

In order to study the dependence on the parameter \fd\ more
extensively, we have run a further set of fits to the data in which we
keep \qrms\ and \n\ fixed at values of 18 \uk\ and 1 respectively, 
and vary \fd\ over the range 0 - 1 in steps of 0.01. In
Fig.~\ref{fig:full_fits_fd} 
the results for the 31.5 GHz channel over the three Galactic cuts
of choice in our analysis are shown. As before, it is clear that 
the derived scaling amplitude decreases as \fd\ increases. This has a
simple physical interpretation, namely that the amount of foreground
emission correlated with the \ha\ template as admitted by the 31.5 GHz
sky map is fixed (see Table~\ref{tab:rms_signal}), and since the
predominant effect of the correction for extinction is simply to
enhance the \ha\ signal with very little change in detailed structure at high
latitudes and over the range of scale probed by $COBE$-DMR, the
coupling amplitude must fall with increasing \fd\ as observed. 
This observation then allows us to restrict our investigations to low
values of \fd\ at high latitudes, consistent with the results of
Dickinson et al. (2003). 

In fact, we can take the analysis yet further. By noting that the scaling amplitude is effectively
related to the value of the electron temperature in the interstellar
medium (\te), then the 95\% confidence limits on the scaling amplitude
imply some joint constraint on \fd\ and \te. By comparison of the predicted scaling
of the \ha\ map, for the temperature range $5000-9000$~K (discussed in
section~\ref{sec:ff_template}), with the 95\% confidence
range for the 31.5 GHz coupling parameters, we infer a weak limit on \fd\
of $\ltsim$~0.75, but this also implies a value for \te\ lower than
the 7000~K observed locally. Indeed, if we adopt this
local value as the typical value for \te\ then \fd\ is constrained to
lower values still, $\ltsim$~0.35, in good agreement with Dickinson
et al. (2003). 

\subsection{Constraints from the 19.2 GHz data}
\label{sec:results_19ghz}

\begin{figure}
  \includegraphics[width=0.5\textwidth]{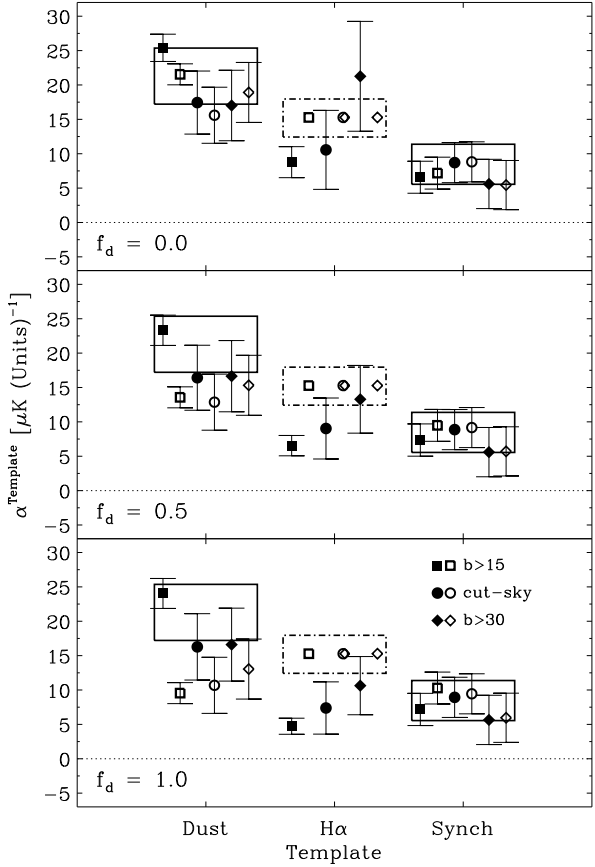}
  \caption{Derived coupling coefficients between the 19.2 GHz data
  and three templates for foreground emission. Nomenclature and
  symbols as in Fig.~\ref{fig:param_fits_fd}.}
  \label{fig:param_fits_19ghz_fd}
\end{figure}

Here we apply the analysis of the previous sections to derive the
correlation coefficients between the three foreground templates and the
19.2 GHz data of Cottingham (1987). This will allow us to
place stronger constraints on the synchrotron and free-free
emissions which rise with decreasing frequency and to investigate any
dust-correlated component at lower frequencies. The results are
summarised in Fig.~\ref{fig:param_fits_19ghz_fd}.
The analysis has again assumed a scale-invariant ($n=1$) CMB
component with normalisation \qrms$=18~\mu$K. We do not find any
significant dependence on the choice of these parameters over the
range allowed by the DMR data\footnote{This is consistent with the analysis of Boughn et
al. (1992) which placed only upper limits on the normalisation
amplitude.}. We also find that the spatially varying noise description is preferred
over the constant noise level per pixel used in the previous analysis
by de Oliveira-Costa et al. (1998). 

Fig.~\ref{fig:param_fits_19ghz_fd} shows that there is a highly
significant correlation between the 19.2 GHz data and all three
foreground templates, particularly for the $|b|>15\degr$ cut.  The free-free coupling
constant calculated as a free parameter agrees most closely with the
expected emission at $f_{d}=0$; high values of $f_{d}$ are
rejected. 

The stronger synchrotron signal at 19.2~GHz compared with that at DMR
frequencies enables a more definitive spectral index between 408~MHz and
these higher frequencies to be obtained as shown in Table 2. The data are
consistent with a brightness temperature spectral index of
$\beta=3.1\pm0.1$ at these frequencies.

The 140~\um\ dust couplings are of
\emph{lower} amplitude than those determined in de Oliveira-Costa et
al. (1998); their analysis did not have the \ha\ template available to
it, and moreover they removed the observed quadrupolar contribution to
the sky signal before the correlation study.  We also repeated our analysis in a similar fashion, using only
the DIRBE 140~\um\ and 408~MHz sky templates,
and we indeed find results more consistent with this earlier
treatment, although approximately $1\sigma$ lower in
amplitude. We also note that de Oliveira-Costa et al. (1998)
found that their statistical error was significantly lower than the
dispersion in coupling amplitudes derived from correlations of the
data with random rotations of the Galactic templates.
The statistical error determined for our analysis is completely consistent with
the value 
determined by such a sky rotation approach. We conclude
that our method is unbiased, and affords statistically robust error estimates.

In the remainder of the paper, we will consider only those results where
we have subtracted the predicted free-free contribution from the
microwave data, and we adopt $f_{d}=0$ and $T_{e}=7000$~K, the values
which are maximally consistent with the picture we have established at
this point.

\section{The nature of the anomalous dust-correlated emission}
\label{sec:results_anomalous}

We have shown that there is a diffuse foreground component in the
$COBE$-DMR data at 31.5 and 53 GHz and the 19.2 GHz data which is
strongly correlated with the 140~\um\ dust emission. In order to make further progress in determining the nature of the
anomalous component of dust-correlated emission, it is necessary to
identify its spectral behaviour over the range of frequencies
studied. This in turn requires that we separate out from the higher
frequency observations that fraction of the observed dust correlation
which is associated with thermal dust emission. 

There are several possible approaches to modelling the thermal dust at DMR
frequencies. 
The first is to use an $IRAS$ or DIRBE template of FIR emission and
extrapolate to microwave frequencies using a simple model of the dust
emissivity and temperature. 
In particular, we could use the model from DL98 where the emissivity index is set
to $\sim~1.7$ and the temperature and optical depth are adjusted so that the peak dust
emission occurs at 140~\um, or set the emissivity index to $\sim$2 and
dust temperature to 17.5~K, as derived in Lagache et al. (2000). 
Kogut et al. (1996a) combined the observed correlations between the
DMR and DIRBE 140 \um\ data with additional information about r.m.s
fluctuations obtained directly from the DIRBE 100, 140 and 240 \um\ 
sky maps to attempt to infer constraints on the thermal
dust emission. These results are quite consistent with the
two previous models. 
Another approach, which we adopt here, is to use a model which
incorporates data available over a wide range of FIR and microwave frequencies;
suitable models have been generated by FDS as described in the next section. 

\subsection{Estimation and removal of thermal dust in the DMR data}
\label{sec:removing_thermal_dust}

\begin{figure}
  \includegraphics[width=0.5\textwidth]{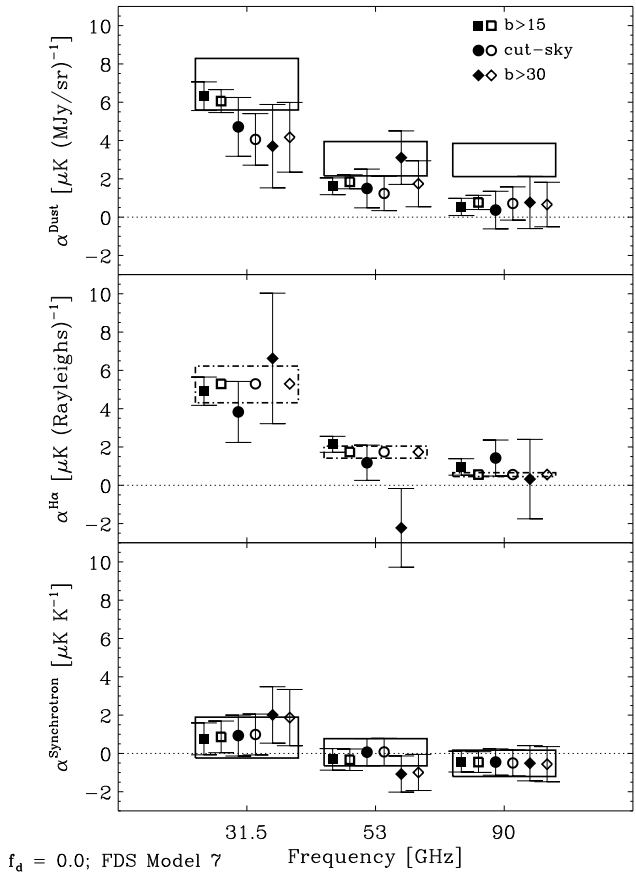}
  \caption{Derived coupling coefficients between the DMR data at
  31.5, 53 and 90 GHz and three templates for foreground emission.
  Nomenclature etc as in Fig.~\ref{fig:param_fits_fd}. Here,
  we remove the contribution from thermal dust
  emission by correcting the DMR data before fitting the templates
  using the predictions and templates from FDS model 7. The free-free
  contribution is estimated only for the template derived assuming
  \fd\ = 0. The results show that the majority of the emission at 90
  GHz is due to thermal dust, with anomalous dust-correlated emission dominating
  at 31.5 GHz, and a mix of the two at 53 GHz.}                                                                                  
  \label{fig:param_fits_noFDS_fd}
\end{figure}

FDS have used $7\degr$ resolution data in 123 frequency lines between
100 and 2100 GHz ($\lambda=3$ mm to $0.14$ mm) from $COBE$-FIRAS data
along with $COBE$-DIRBE data in the range 100 to 240 \um\ to produce
models of dust emission which can be extrapolated to DMR
frequencies. Their models 7 and 8 afford the
most acceptable fits to the observations, according to a chi-square
criterion. We have adopted model 7, differing little from model 8,
which has two dust components, one at 9.6~K with $\alpha=1.5$ and the
other at 16.4~K with $\alpha=2.6$. We proceed by subtracting from the 
DMR data the model 7 predictions at each frequency, then
correlate the residuals against the DIRBE and Haslam templates as
before, with either an unconstrained and fitted free-free
contribution, or after subtracting the predicted
free-free component assuming \fd\ = 0 and \te = 7000~K. 
Fig.~\ref{fig:param_fits_noFDS_fd} shows the results from this
analysis. The conclusions reached based on model 8
are indistinguishable from those presented here (although model
7 has $\sim\ 20\%$ higher r.m.s fluctuations at 90 GHz relative
to model 8).

It is apparent when comparing Fig.~\ref{fig:param_fits_noFDS_fd} with
the left-hand panel of Fig.~\ref{fig:param_fits_fd}, that the thermal dust emission accounts for 
essentially the entire observed correlated signal at 90 GHz. The best
estimate of the dust components comes from the $|b|>15\degr$ cut and
fixed \ha\ data. Here we find that at 90 GHz the model thermal dust
component is equivalent to $1.39$~\uk/(MJy~sr$^{-1}$) at 140~\um\
leaving $0.77 \pm 0.37$~\uk/(MJy~sr$^{-1}$) of anomalous dust-correlated emission. This signal
may be indicative of a small anomalous component, but equally it could
be a consequence of the FDS parameterisation inadequately 
predicting an enhanced (very) cold dust component (for example, see Reach et
al. 1995). However, with the current data it is premature to radically alter the FDS
model.

The most important result of this study is the high statistical
significance of the identified anomalous dust-correlated component of $6.06 \pm 0.60$
\uk/MJy~sr$^{-1}$ at 31.5 GHz. At 53 GHz the dust-correlated DMR
coefficient is $1.85 \pm 0.36$ \uk/(MJy~sr$^{-1}$), of which the thermal
dust model 7 contributes $0.57$ \uk/(MJy~sr$^{-1}$). The two
dust-correlated components will therefore be approximately equal in
amplitude at frequencies of $\sim 60$ GHz and suggests that this may
be close to the minimum in terms of Galactic foreground contamination
on large ($\sim 7^{\circ}$) angular scales. The dust contribution is
discussed further in sections \ref{sec:wmap_comparison} \& \ref{sec:conclusions}.   

\subsection{Characterising the anomalous dust-correlated spectrum}
\label{sec:anom_dust_spectrum}

In order to quantify the nature of the anomalous dust-correlated component, we
first compute the effective power-law spectral index between pairs of
frequencies, as tabulated in Table~\ref{tab:anomalous_index}. 
We use the 19.2 GHz results directly from
Fig.~\ref{fig:param_fits_19ghz_fd}, 
without further correction for the thermal dust contribution since
this is negligible at this frequency (comprising $\sim\ 50\%$ of
the thermal contribution at 31.5 GHz which is in itself negligible).
In addition, all data points are fitted simultaneously by a model of
the form ${\mathbf A_{anom}\, \left(\frac{\nu}{\nu_0}\right)^{-\beta}}$
where ${\mathbf A_{anom}}$, in units of \uk/(MJy~sr$^{-1}$) referred
to the 140~\um\ template, is the amplitude of the emission at
the reference frequency $\nu_0$, which we take as 31.5 GHz.
The values are consistent with a spectrum steeper than the typical
value $2.15$ of free-free emission, 
although in themselves not ruling out such an emission mechanism. 
Previous interpretations had proposed free-free as the source of the
excess dust-correlated emission, yet in this work we have utilised a sky map of
\ha\ optical line emission as an explicit tracer of
this foreground component. The derived correlation with the data is
consistent with that predicted for free-free emission and moreover,
does not eliminate the dust-correlated emission.

\begin{table}
  \begin{center}
  \caption[Anomalous dust spectral index]{The effective spectral index
           between pairs of frequencies for the anomalous dust-correlated
           component (after correction for free-free emission
           assuming $f_d = 0$ and for thermal dust emission using
           model 7 of FDS). 
           The bottom line represents the best fit
           spectral index for a power law model
           of form ${\mathbf A_{anom}}\, \left(\frac{\nu}{\nu_0}\right)^{-\beta}$ where $\nu_0$ is
           taken as 31.5 GHz computed over all frequencies. 68\%
           c.l. errors are reported.}
  \label{tab:anomalous_index}   
  \begin{tabular}{cccc} \hline
                    & \multicolumn{3}{c}{Sky Coverage} \\ \cline{2-4}
                    & $|b| > 15\degr$   & Standard        & $|b| > 30\degr$   \\ \hline\hline
  $\beta_{19:31}$  & 2.51$\pm$ 0.30  & 2.66$\pm$ 0.78  & 2.99$\pm$ 0.80    \\ 
  $\beta_{31:53}$  & 2.28$\pm$ 0.66  & 2.28$\pm$ 1.60  & 1.68$\pm$ 1.47   \\ 
  $\beta_{53:90}$  & 1.66$\pm$ 1.54  & 1.04$\pm$ 2.82  & 1.84$\pm$ 3.36  \\ \hline
   Fitted $\beta$  & 2.44$^{+0.29}_{-0.26}$ & $2.51^{+1.32}_{-0.94}$ & $2.65^{+1.77}_{-1.01}$ \\ \hline
   Fitted ${\mathbf A_{anom}}$ & 6.34$^{+0.69}_{-0.72}$ & 4.35$^{+1.63}_{-1.87}$ & 4.93$^{+2.22}_{-2.74}$ \\ \hline
  \end{tabular}
  \end{center}
\end{table}

\begin{figure}
  \includegraphics[width=0.5\textwidth]{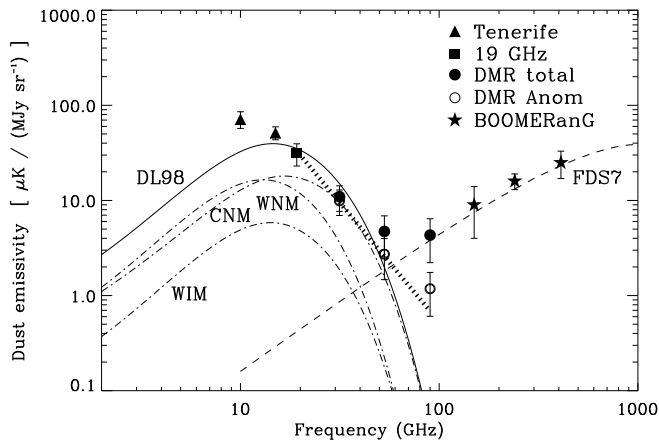}
  \caption{Dust emissivity correlation coefficients referred to 100
  \um, in units
  of \uk/(MJy~sr$^{-1}$) for $|b| \gtsim 20\degr$. Data points are: Tenerife at 10 and 15 GHz
  (filled triangles; de Oliveira-Costa et al. 2002), 19.2 GHz (filled
  square; this work), DMR at 31.5, 53 and 90 GHz (filled circles; this
  work) and BOOMERanG at 150, 240, and 410 GHz (filled stars; Masi et
  al. 2001). Also plotted are the DMR correlations $after$ the thermal
  dust component has been subtracted (open circles) using model FDS7
  (dashed line). The best fitting power-law model for the anomalous
  component is shown as a heavy dotted line over the range $19-90$
  GHz. Spinning dust models from DL98 for dust in 3 phases of the
  interstellar medium (CNM, WNM, WIM) are shown as dot-dashed lines scaled
  to 100~\um\ (see text). The preferred total spinning dust model from
  DL98 is shown as a solid line.}
  \label{fig:dust}
\end{figure}

Microwave data on the anomalous dust-correlated emission in addition to
that given in the DMR and the 19.2 GHz surveys made at intermediate
Galactic latitude ($|b| \gtsim 20\degr$), are presented in
Fig.~\ref{fig:dust}. To allow a direct comparison with these
previous studies, we have recomputed the DMR and 19.2 GHz correlation coefficients at
100~\um. These are a factor of $\sim 2$ higher than the corresponding
values for the 140 \um\ correlations, reflecting the weaker dust
emission at 100 \um. Both the total dust
(filled circles) and anomalous dust-correlated (open circles) emission
are shown for
comparison. The best-fit power law is shown over the frequency range
$19-100$ GHz. Tenerife data on a
5\degr scale at 10 and 15 GHz give correlation coefficients of $71.4 \pm
14.2$ and $51.4 \pm 8.0$ \uk/(MJy~sr$^{-1}$) for $|b|>20\degr$ at 100~\um\ respectively
(de Oliveira-Costa et al. 2002). Mukherjee
et al. (2001) suggested lower values based on a smaller Tenerife
database. Finkbeiner et al. (2002) have also studied two isolated dust
clouds on the Galactic plane at 5,
8.25 and 9.75 GHz; the dust emissivity differs by more than a factor
of 10 in the two clouds and hence they have not been
plotted. Nevertheless they find a dust-correlated emissivity which
falls between 10 and 5 GHz. There is
a suggestion of a turn-over in the anomalous component spectrum at $\sim
10$ GHz but the exact shape is still poorly defined. Information at
frequencies of $5-15$~GHz is clearly
important in defining the anomalous dust-correlated spectrum. A large area survey
at intermediate and high Galactic latitudes has been made with the
Jodrell Bank Interferometer at angular scales of 1\degr and 2\degr
(Jones et al. 2001). At 2\degr resolution the observed r.m.s level is
$86 \pm 12$ \uk; this includes all the foreground components and the
CMB. The synchrotron and free-free components comprise more than half
this value, leaving $\ltsim 50$ \uk\ for dust emission. The
corresponding dust emissivity at 5 GHz is $\ltsim 100$ \uk/(MJy
sr$^{-1}$) at 140~\um. 


Fig.~\ref{fig:dust} also shows dust data at 150, 240 and 410 GHz from the BOOMERanG experiment
(Masi et al. 2002) also for $|b|>20\degr$, corrected from
thermodynamic units to brightness temperature units. The FDS thermal dust model
7 normalised to 100~\um\ (dashed line) is in relatively good agreement
with these points data. 

Fig.~\ref{fig:dust} therefore presents our assessment of the relative
contribution of the anomalous and thermal components of dust emission
in the range 10 to 500 GHz. The contributions are equal at $\sim 60$
GHz. These results apply to large areas of sky ($\gtsim 1$~sr) at
intermediate and high latitudes where deep CMB observations are made;
the situation closer to the plane near sources of heating and
star-formation is likely to be different.

The best-fitting model to the anomalous dust-correlated component
is a power law model spectrum of the form $\nu^{-\beta}$ with $\beta
\sim 2.5$, in the range $19-90$~GHz. This implies an emission
component with a dust-like morphology but a synchrotron-like
spectrum. The physical
explanation of such a component is open to debate, and we explore
dust-related alternatives below. It should be noted, however, that the
evidence presented here for dust-correlated emission does not
necessarily imply that the emission is intrinsic to the dust grains
themselves. For example, it may arise from a source which is
associated with the dust, such as may result from the star-formation
process in a spiral arm environment. This is the interpretation taken
by the $WMAP$ team (Bennett et al. 2003) and is discussed in section
\ref{sec:wmap_comparison}. 

\subsection{Possible models of the anomalous emission}
\label{sec:alternative_models}

A widely discussed model is in terms of rotational emission from
very small dust grains -- a \lq spinning' dust 
component, as proposed in DL98. They propose a three-component
model of the spinning dust containing contributions from dust found
in the Cold Neutral Medium (CNM), Warm Neutral Medium (WNM) or Warm
Ionised Medium (WIM). The spectra for these
three environments with the inferred anomalous emission are shown in
Fig.~\ref{fig:dust} scaled to 100~\um\ as dot-dashed lines. The
preferred DL98 model has these components in the ratio 0.43, 0.43 and
0.14 respectively. The typical spectral index of the calculated dust emission over the
frequency range $19 - 53$~GHz is between $3.3$ and $4$, reasonably
consistent with the observed values (Table \ref{tab:anomalous_index}). 
However, the results presented in this paper indicate that,
whilst the preferred spinning dust model of DL98 has a spectral
{\em shape} which can reasonably match the data, some adjustment of
amplitude is required. Fig.~\ref{fig:dust} shows
the model (solid line) scaled to approximately fit the
plotted data points. A more formal $\chi^2$ analysis indicates that
an improved fit results from models in which emission from the WNM 
is almost completely dominant. This can be understood rather easily:
the CNM-component can be adjusted to give an excellent fit to the
data points at 19.2 and 53 GHz, but then over-predicts emission
at 31.5 GHz; the WNM-component is scalable to provide good
agreement with the data at 19.2 and 31.5 GHz although then
under-predicts the emission at 53 GHz; the WIM-component
can only be matched well to one frequency. Since the two lowest
frequency data points have the most statistical weight, models
containing a substantial contribution from the WNM are favoured.
Spinning dust models therefore remain a possibility to explain
the anomalous dust-correlated component, although the observed
spectrum is not as strongly peaked as the current dust models 
which therefore require some adjustments to match more closely with the new constraints.
Given the uncertainties in the 
model predictions, depending on the composition, size and shape 
of the dust grains and conditions in the interstellar medium, 
this is certainly feasible.

An alternative model for the anomalous dust-correlated emission 
at frequencies $\ltsim 100$~GHz is due to thermal fluctuations in
the magnetization of interstellar grains (Draine \& Lazarian 1999)
which results in magnetic dipole radiation. A
population of Fe grains or grains with Fe inclusions are predicted to
produce strong emission with a broad peak near 70~GHz due to the
Fr\"{o}hlic resonance. Normal paramagnetic grains give much weaker
emission at a level of $10^{-2}$ of the dipole emission expected from
spinning dust grains. Hypothetical Fe-rich materials could produce
radiation comparable to that from spinning grains in the 14 to 19~GHz
range. Draine \& Lazarian (1999) conclude that magnetic dipole
emission at $\nu < 100$~GHz might contribute significant emission if appropriate magnetic materials existed;
both astronomical and laboratory measurements could clarify this
situation. In the current context, the spectrum of the radiation is
quite different; the magnetic dipole emission is relatively flat while
the dipole (spinning) emission is peaked.

A possible explanation for excess microwave emission in the North
Celestial Pole region, at 14.5 and 31.7
GHz, was proposed by Leitch et al. (1997) in terms of free-free
emission from ionized gas at $T_{e} > 10^{6}$~K. This explanation was considered unlikely because of
the energetics in creating this hot gas. Draine \& Lazarian (1998a)
argue that the case against this hypothesis is even stronger for the larger DMR sky
coverage since the power radiated is a factor of $\sim 100$ greater
than that expected from the typical supernova rate in the Galaxy. The
current analysis also argues against this in view of the strong dust correlation and the
fact that it persists in the presence of a simultaneous fit to the
\ha\ template. However, it should be noted that the ratio of \ha\ to
free-free radio continuum drops as the gas temperature is increased
above $\sim 3 \times 10^{4}$~K, hence the requirement for temperatures
above $\gtsim 10^{6}$~K. The definitive test of this model would
be to analyse the correlation of microwave data with X-ray data at
energies of $\sim 1$~keV. Indeed Kneissl et al. (1997) did find a
significant ($>2\sigma$) correlation of DMR data with ROSAT all-sky X-ray data but
only for the quadrupole $\ell=2$ component. After removal of the quadrupole term, no
significant correlation remained. 

Another possible source of microwave emission from dust grains is in
terms of low-lying 2-level tunnelling transitions at low
temperatures. The laboratory measurements of Agladze et al. (1996)
show surprisingly large and variable values for the millimetre (80 to
400~GHz) absorption coefficients of small silicate particles similar
to interstellar grains; these effects occur at temperatures found in
the interstellar medium (T $\ltsim 20$~K). This phenomenon is also
found in amorphous silica-based glasses and is believed to be due to
resonant tunnelling absorption, although the detailed physics is not
fully understood (B\"{o}sch 1978, Agladze \& Sievers 1998). The low
frequency spectrum of this emission has not yet been determined so a
comparison with Fig.~\ref{fig:dust} is not possible. Further progress awaits
laboratory measurements.

\subsection{Comparison with {\it WMAP}}
\label{sec:wmap_comparison}

Since our more detailed re-analysis of the $COBE$-DMR data, the first results
from the analysis of one year of observations by the {\it WMAP}
satellite have been released (Bennett et al. 2003a and references therein). Their
foregrounds analysis (Bennett et al. 2003b) uses a maximum entropy
method applied to the 5 frequency channels of the {\it WMAP}
observations, together with data from outside the microwave spectral
regime, specifically the Haslam 408 MHz survey, the full-sky \ha\
map published by Finkbeiner (2003) and a dust template derived
from $COBE$-DIRBE and FIRAS data (FDS model 8 normalised at 94 GHz). 
These external data-sets are claimed to be
the key to separating the three emission mechanisms 
(free-free, synchrotron and spinning dust) with a spectral
index of $\beta \sim 2$ at the {\it WMAP} frequencies. One should
note, however, that no unique spinning dust template has been employed
since this putative emission falls off rapidly outside the microwave
regime, and therefore there is no obvious candidate (although DL98
have proposed that the $DIRBE$ 12 \um\ map might be used).
More importantly, the 
analysis is allowed only to produce a combined synchrotron/spinning dust solution
at each frequency, with no attempt made to separately disentangle
these two components. As a result of the analysis, they identify a
dust-correlated component in the lower frequency (23, 33 and
41 GHz) {\it WMAP} channels. However, they interpret this as a
synchrotron component with a spectral index of $\beta \sim 2.5$,
a value which we emphasise is 
in close agreement to that derived in this analysis using $COBE$-DMR
data. They argue that this
emission originates in star-forming regions close to the plane
while the steeper ($\beta \sim 3)$ synchrotron component is from gas
in the halo, a picture which is consistent with what has been
found in the integrated radio and FIR emission for several nearby star-forming
galaxies. 

In their figure 9, Bennett et al. (2003b) show the steepening of the
mode for the spectrum of the synchrotron plus spinning dust component
at 23-41 GHz and conclude that this is evidence for a
spectral break, corresponding to a steepening of $\Delta \beta \sim 0.5$ at
frequencies of $\sim 20$ GHz. However, this could also be the
signature of spinning dust. Current models of spinning dust (DL98) also give
a steepening of index with spectral indices ranging from $\sim 2.5$ at
30 GHz and steepening to greater than 4 at frequencies $\gtsim 50$ GHz. 

An independent re-analysis of the {\it WMAP} foregrounds has been
performed by Lagache (2003). {\it WMAP} data is correlated with the neutral H{\sc i}
component and an excess is similarly found at longer wavelengths. By
comparing the correlations with the H{\sc i} column density, 
there is evidence that the excess emission 
is associated with small transiently heated
dust particles. The exact physical mechanism of
such emission remains to be determined,  
but possible models certainly include the spinning dust rotational
emission mechanism or Very Small Grain (VSG) long-wavelength emission
resulting from transient heating by UV photons.

In the previous discussion, we have attempted to present
arguments as to why the interpretation of the {\it WMAP} foreground
results remains open, and that
a spinning dust component can be compatible with the
observations. However, we would like to emphasise that we
do {\em not} consider that this is an unambiguous
solution to the origin of the dust-correlated emission.
The results of our analysis based on independent data sets
are completely consistent with that of the {\it WMAP} team, and indeed the
flat spectrum fit we have derived to the anomalous dust-correlated
component is formally better than any spinning dust model considered.
In order to clearly distinguish the various models for this excess
emission, low frequency data (few GHz to 15 GHz) will be
vital. Evidence for {\it WMAP}-consistent spectral indices determined
from lower frequency radio measurements may be important to
corroborate the star-forming region, flat spectrum hypothesis.
We await the forthcoming publication of a full-sky 1420 MHz survey
with interest. 
Finally, detailed spatial comparisons between the radio and FIR
emission will be critical in determining the physical processes
associated with this dust-correlated component.



\section{Results - CMB}
\label{sec:results_CMB}

We are now in a position to consider the implications of our better
understanding of the Galactic diffuse foregrounds upon estimates of
\qrms, the CMB quadrupole normalisation and $n$, the power-law
spectral index. In section~\ref{sec:method} we describe how the
simultaneous solutions for the coupling coefficients of the DMR data
to the three foregrounds forces the CMB distributed power to be
invariant between the three DMR frequencies, so that each set of
foreground solutions has an accompanying estimate of \qrms\ and
$n$. Table \ref{tab:qn_fits} gives the \qrms\ and $n$ values for the
following areas of parameter space.
\begin{itemize}
\item{Column 1: Foregrounds used in the fit-- 140~\um\ dust (D),
synchrotron (S), free-free (F), free-free forced to the \ha-derived
value (FixedF) and a fixed thermal dust component following FDS model 7
(FDS7).}
\item{Column 2. The Galactic cuts used in the fits-- straight Galactic
latitude cuts at $|b|>15\degr$ and $30\degr$, and the standard DMR 
cut at $|b|>20\degr$ which also excludes extra regions of stronger emission.}
\item{$f_{d}$, the correction applied to the free-free (\ha) template
for foreground dust absorption. We consider $f_{d}=0.0$ is the best
value for the latitudes considered here (see section~\ref{sec:results_dmr_ff}).}
\end{itemize}

An examination of Table \ref{tab:qn_fits} immediately shows that the
solution for \qrms\ and $n$ is strongly independent of the parameters
incorporated in the present study. We believe that our best estimate
of \qrms\ and $n$ are given for $f_{d}=0$, fixed free-free and the FDS
dust model included, namely \qrms$=15.18^{+2.81}_{-2.30}$ \uk,
$n=1.23^{+0.22}_{-0.23}$. This may be compared with
\qrms$=16.49^{+3.13}_{-2.49}$ \uk, $n=1.19^{+0.22}_{-0.23}$ if only
the dust and synchrotron foregrounds are used with an aggressive
Galactic cut; the earlier analysis of
Bennett et al. (1996) gave \qrms$=15.3^{+3.8}_{-2.8}$ \uk, $n=1.2 
\pm 0.3$. The present analysis gives values in excellent agreement with the
earlier values, but with smaller errors.

The reason that the solutions for \qrms\ and $n$ in Table
\ref{tab:qn_fits} are not strongly dependent upon the precise values
of the Galactic foregrounds used is that the weighting of the data
from the three radiometers with respect to the likelihood analysis
is strongest for 53 and 90~GHz; the weightings are approximately 
8, 65, and 27 per cent at 31.5, 53 and 90
GHz respectively, which reflects the corresponding noise levels of
the three frequencies. As will be shown in section~\ref{sec:conclusions},
the minimal foreground contributions are in the range 50 to 100~GHz and
therefore explains why there is only a small effect on \qrms\ and $n$.


Indeed, even for a Galactic cut of 15\degr, there
appears to be remarkable consistent with previous analyses, 
provided the three templates for foreground emission are
included in the analysis. 
The worst perturbation 
in \qrms\ and \n\ 
corresponds to the result from the data corrected for
dust and synchrotron alone for this cut. However,
even here the changes are less than  1$\sigma$ for the quadrupole
normalisation, and a less significant amount for 
the cosmological spectral index value.
Clearly, the uncorrected fit 
for this latitude cut is inconsistent at worse than 
2$\sigma$. 

\begin{table*}
  \begin{center}
  \caption[]{Cosmological parameter fits to the DMR data for a power
  law model parameterised by r.m.s quadrupole normalisation amplitude
  \qrms\ (\uk) and power law index \n. The fits are made for three Galactic
  cuts, and after correction for various combinations of foreground
  emission as determined by simultaneous fits to template sky
  maps. The corrections should be understood as follows: D - dust
  traced by the DIRBE 140~\um\ sky map; F - free-free traced by the
  \ha\ template of Dickinson et al. (2003) after correction for 
  dust absorption specified by \fd\ = 0, 0.5 and 1.0; S - synchrotron emission
  determined from the 408~MHz survey of Haslam et al. (1982); FixedF -
  we subtract a fixed free-free contribution determined by scaling the \ha\
  template using the amplitudes predicted in Dickinson et al. for an
  electron temperature of 7000~K after correcting for dust absorption specified
  by \fd\ = 0, 0.5 and 1.0;
  FDS7 - we subtract a thermal dust contribution as predicted by FDS
  Model 7. }
  \label{tab:qn_fits}
  \begin{tabular}{llcccccc} \hline

    Correction & Coverage & \multicolumn{3}{c}{$Q_{rms-PS}$} & \multicolumn{3}{c}{$n$} \\ \cline{3-8}
               & & $f_d\, = $ 0.0 & 0.5 & 1.0 & $f_d\, = $0.0 & 0.5 & 1.0 \\ \hline\hline
                 & $b>15\degr$ & 24.45$^{+ 4.60}_{-3.68}$&  - &  - &  0.94$^{+ 0.21}_{-0.22}$&  - &  -  \\
     Uncorrected &    standard & 15.85$^{+ 3.26}_{-2.59}$&  - &  - &  1.22$^{+ 0.24}_{-0.26}$&  - &  -  \\
                 & $b>30\degr$ & 15.31$^{+ 3.55}_{-2.78}$&  - &  - &  1.20$^{+ 0.28}_{-0.29}$&  - &  -  \\ \hline
                 & $b>15\degr$ & 16.49$^{+ 3.13}_{-2.49}$&  - &  - &  1.19$^{+ 0.22}_{-0.23}$&  - &  -  \\
              DS &    standard & 15.08$^{+ 3.04}_{-2.43}$&  - &  - &  1.23$^{+ 0.24}_{-0.25}$&  - &  -  \\
                 & $b>30\degr$ & 15.02$^{+ 3.39}_{-2.65}$&  - &  - &  1.19$^{+ 0.27}_{-0.29}$&  - &  -  \\ \hline
                 & $b>15\degr$ & 15.34$^{+ 2.91}_{-2.30}$& 15.18$^{+ 2.84}_{-2.30}$& 15.12$^{+ 2.81}_{-2.30}$&  1.22$^{+ 0.22}_{-0.24}$&  1.23$^{+ 0.22}_{-0.23}$&  1.23$^{+ 0.22}_{-0.23}$ \\
             DSF &    standard & 15.31$^{+ 3.13}_{-2.46}$& 15.37$^{+ 3.10}_{-2.49}$& 15.34$^{+ 3.10}_{-2.46}$&  1.21$^{+ 0.24}_{-0.25}$&  1.20$^{+ 0.24}_{-0.25}$&  1.20$^{+ 0.24}_{-0.25}$ \\
                 & $b>30\degr$ & 14.70$^{+ 3.39}_{-2.62}$& 15.18$^{+ 3.42}_{-2.72}$& 15.18$^{+ 3.45}_{-2.72}$&  1.22$^{+ 0.27}_{-0.29}$&  1.18$^{+ 0.27}_{-0.29}$&  1.18$^{+ 0.27}_{-0.29}$ \\ \hline
                 & $b>15\degr$ & 15.34$^{+ 2.91}_{-2.33}$& 15.24$^{+ 2.91}_{-2.30}$& 15.34$^{+ 2.91}_{-2.33}$&  1.22$^{+ 0.22}_{-0.24}$&  1.22$^{+ 0.22}_{-0.24}$&  1.23$^{+ 0.22}_{-0.24}$ \\
       DS-FixedF &    standard & 15.47$^{+ 3.13}_{-2.46}$& 15.63$^{+ 3.16}_{-2.49}$& 15.79$^{+ 3.20}_{-2.52}$&  1.20$^{+ 0.24}_{-0.25}$&  1.18$^{+ 0.24}_{-0.25}$&  1.17$^{+ 0.24}_{-0.25}$ \\
                 & $b>30\degr$ & 15.66$^{+ 3.52}_{-2.78}$& 15.56$^{+ 3.48}_{-2.78}$& 15.66$^{+ 3.52}_{-2.78}$&  1.15$^{+ 0.27}_{-0.28}$&  1.15$^{+ 0.27}_{-0.29}$&  1.15$^{+ 0.27}_{-0.29}$ \\ \hline
                 & $b>15\degr$ & 15.12$^{+ 2.84}_{-2.30}$& 14.86$^{+ 2.75}_{-2.24}$& 14.80$^{+ 2.78}_{-2.24}$&  1.23$^{+ 0.22}_{-0.23}$&  1.25$^{+ 0.22}_{-0.23}$&  1.26$^{+ 0.22}_{-0.23}$ \\
         DS-FDS7 &    standard & 15.18$^{+ 3.07}_{-2.46}$& 15.18$^{+ 3.07}_{-2.43}$& 15.18$^{+ 3.04}_{-2.46}$&  1.21$^{+ 0.24}_{-0.25}$&  1.21$^{+ 0.24}_{-0.25}$&  1.21$^{+ 0.24}_{-0.25}$ \\
                 & $b>30\degr$ & 14.64$^{+ 3.36}_{-2.62}$& 15.05$^{+ 3.42}_{-2.68}$& 15.05$^{+ 3.45}_{-2.65}$&  1.22$^{+ 0.27}_{-0.29}$&  1.19$^{+ 0.27}_{-0.29}$&  1.19$^{+ 0.27}_{-0.29}$ \\ \hline
                 & $b>15\degr$ & 15.18$^{+ 2.81}_{-2.30}$& 14.86$^{+ 2.75}_{-2.27}$& 14.83$^{+ 2.81}_{-2.24}$&  1.23$^{+ 0.22}_{-0.23}$&  1.25$^{+ 0.22}_{-0.23}$&  1.27$^{+ 0.22}_{-0.24}$ \\
  DS-FixedF-FDS7 &    standard & 15.28$^{+ 3.07}_{-2.43}$& 15.37$^{+ 3.13}_{-2.43}$& 15.56$^{+ 3.10}_{-2.52}$&  1.20$^{+ 0.24}_{-0.25}$&  1.20$^{+ 0.24}_{-0.25}$&  1.18$^{+ 0.24}_{-0.25}$ \\
                 & $b>30\degr$ & 15.50$^{+ 3.48}_{-2.72}$& 15.43$^{+ 3.48}_{-2.75}$& 15.53$^{+ 3.52}_{-2.75}$&  1.16$^{+ 0.27}_{-0.29}$&  1.16$^{+ 0.27}_{-0.29}$&  1.15$^{+ 0.27}_{-0.29}$ \\ \hline

  \end{tabular}
  \end{center}
\end{table*}

Given the success of the Galactic correction, we might wonder
whether it is possible to be yet more aggressive, and attempt to
correct for Galactic contamination at still lower latitudes. We have
tried this using a cut in which we only eliminate those pixels lying within
10\degr of the Galactic plane, but in this case we find that, whilst
the derived correlations of the foreground templates with the DMR maps
are very consistent with the 15\degr results discussed above, the
likelihood analysis begins to deteriorate. This close to the Galactic plane,
there remains a sufficiently complex structure in the Galactic
emission that it is inevitable that simple correlation studies which
do not allow for variations in spectral behaviour cannot succeed in
subtracting the foreground at the necessary level of
accuracy. Moreover, for this case, half of the $7\degr$ FWHM beam area of DMR lies at
$|b|<10\degr$. Residual poorly corrected features contribute to an overall flattening
of the cosmological power-law spectral index and increase in
normalisation amplitude well beyond the allowed ranges yielded by the
studies confined to more conventional latitude cuts. It is unlikely that even
sophisticated analysis methods, such as those based on independent
component analysis (Maino et al. 2002) or Maximum Entropy in harmonic 
space (Stolyarov et al. 2002) will
be able to circumvent these issues. With higher sensitivity, higher
angular resolution observations such as those due to {\it WMAP}
and expected for {\it Planck}, such methods as applied to small regions
or even on a pixel-by-pixel basis may yield an improved 
understanding of the physical nature of the foreground emission
and allow greater sky coverage to be used for cosmological studies.


\section{Conclusions}
\label{sec:conclusions}

In this paper we have reappraised the contribution of Galactic
foreground emission to the $COBE$-DMR data, augmented with lower
frequency information at 19.2~GHz. It has been demonstrated that a new
template for Galactic free-free emission derived from observations of
the \ha\ emission is in excellent agreement with expectations, and
furthermore allows us to claim, with reasonable certainty, that the
amount of emission absorbed by foreground dust is low at high Galactic
latitudes, in particular for the preferred local value of the electron
temperature of 7000~K. After subtraction of the free-free contribution,
there is continued and unambiguous evidence of anomalous dust-correlated emission at low
frequencies, consistent with previous analyses. Subtraction of an FDS model for
Galactic thermal dust emission indicates that the correlated dust
component at 90~GHz is essentially dominated by a vibrational (thermal) emission
mechanism. 
The anomalous component at lower frequencies ($19-90$~GHz) has a power-law spectral
index of $\sim 2.5$, somewhat flatter than the putative spinning dust
component proposed in DL98. Indeed, their preferred mixture of
spinning dust emission from a three-component dust model 
requires some adaptation to afford an improved fit. 
The level of Galactic synchrotron
emission is consistent with expectations based on lower frequency
observations at 408~MHz and a steep temperature spectral index of
$\sim 3.1$. The best-fit coupling coefficients ($|b|>15$\degr) for each foreground
component are given in
Table \ref{tab:summary_table} for 19.2, 31.5, 53 and 90~GHz. Both the
total and anomalous dust-correlated are given for comparison while the
free-free scalings are those predicted by the \ha\ template assuming
$T_{e}=7000$~K and $f_{d}=0$. These results are completely consistent with the
recent {\it WMAP} analysis and independently confirm their
results on large angular scales. In particular, we demonstrate the existence of an
emission mechanism that has a spatial morphology similar to dust but
with a synchrotron-like spectrum in the range $19-90$ GHz.

\begin{table*}
  \begin{center}
 \caption[]{Definitive Galactic Template Coupling Coefficients: 
             derived assuming a fixed free-free contribution
             with \te\ = 7000~K and $f_{d}=0$ for the Galactic latitude
             range $|b|>15\degr$. The FDS7 model for thermal dust emission
             has been applied in order to allow the anomalous
             dust-correlated emission contribution to be unambiguously determined.
             The Galactic coefficients have units $\mu{\rm K}~X^{-1}$ where $X$ are the
             natural units of the template map: MJy~sr$^{-1}$ for the 140~\um\
             DIRBE dust map, Kelvin for the 408~MHz synchrotron map and Rayleigh for the
             \ha\ (free-free) map. The free-free scaling coefficients 
             are determined according to the relations in Dickinson
             et al. (2003), assuming $T_{e}=7000$~K for the central
             value, and 5000 and 9000~K for the lower and upper
             error bars respectively.}
  \begin{tabular}{lcccc} \hline
    Galactic    & \multicolumn{4}{c}{Frequency (GHz)} \\ \cline{2-5}
    Coefficient [units]& 19.2 & 31.5 & 53 & 90 \\ \hline \hline
Total Dust $[\mu$K/MJy sr$^{-1}]$  &  21.55 $\pm$   1.53 &   6.25 $\pm$   0.60 &   2.42 $\pm$   0.36 &   2.16 $\pm$   0.37 \\ 
Anom. Dust $[\mu$K/MJy sr$^{-1}]$  &  21.55 $\pm$   1.53 &   6.06 $\pm$   0.60 &   1.85 $\pm$   0.36 &   0.77 $\pm$   0.37 \\ 
Synchrotron $[\mu$K/K$]$ &   7.18 $\pm$   2.32 &   0.91 $\pm$   0.83 &  -0.26 $\pm$   0.56 &  -0.35 $\pm$   0.54 \\ 
Free-free $[\mu$K/$R]$   &  15.28$^{+2.68}_{-2.85}$ &   5.30$^{+0.93}_{-0.99}$ &   1.74$^{+0.30}_{-0.32}$ &   0.56$^{+0.10}_{-0.10}$ \\ \hline 
\label{tab:summary_table}  
 \end{tabular}
  \end{center}
\end{table*}

Our better understanding of the foregrounds allows us to correct for
them over a larger region of the sky and reassess the value of
\qrms\ and $n$. 
The most significant
changes in general result from an analysis made with an aggressive $|b|>15\degr$
Galactic cut, yet even here the results are compatible with earlier
results at better than the $\sim\ 1\sigma$ level, although the error is slightly
smaller, suggesting some possibility 
of making use of modestly larger sky coverage for the inference 
of cosmological parameters. Interestingly, what might be considered
our best model for the foreground correction -- with this aggressive
Galactic cut, utilising the FDS Model 7
dust template to correct for thermal dust emission, the \ha\ template
from Dickinson et al. (2003) to trace the free-free emission (fixed assuming
that the electron temperature is 7000~K), together with unrestricted
fits to the DIRBE 140~\um\ and 408~MHz sky surveys to correct for
anomalous dust-correlated and synchrotron emission respectively -- results in 
shifts in \qrms\ and \n\ of only $\sim 0.6$~\uk\ and $\sim 0.01$
relative to the previous standard uncorrected result.
All results with $f_{d}=0$ are consistent with values of \qrms\ =
15.3 \uk\ and \n\ = 1.2.

\begin{figure}
  \includegraphics[width=0.5\textwidth]{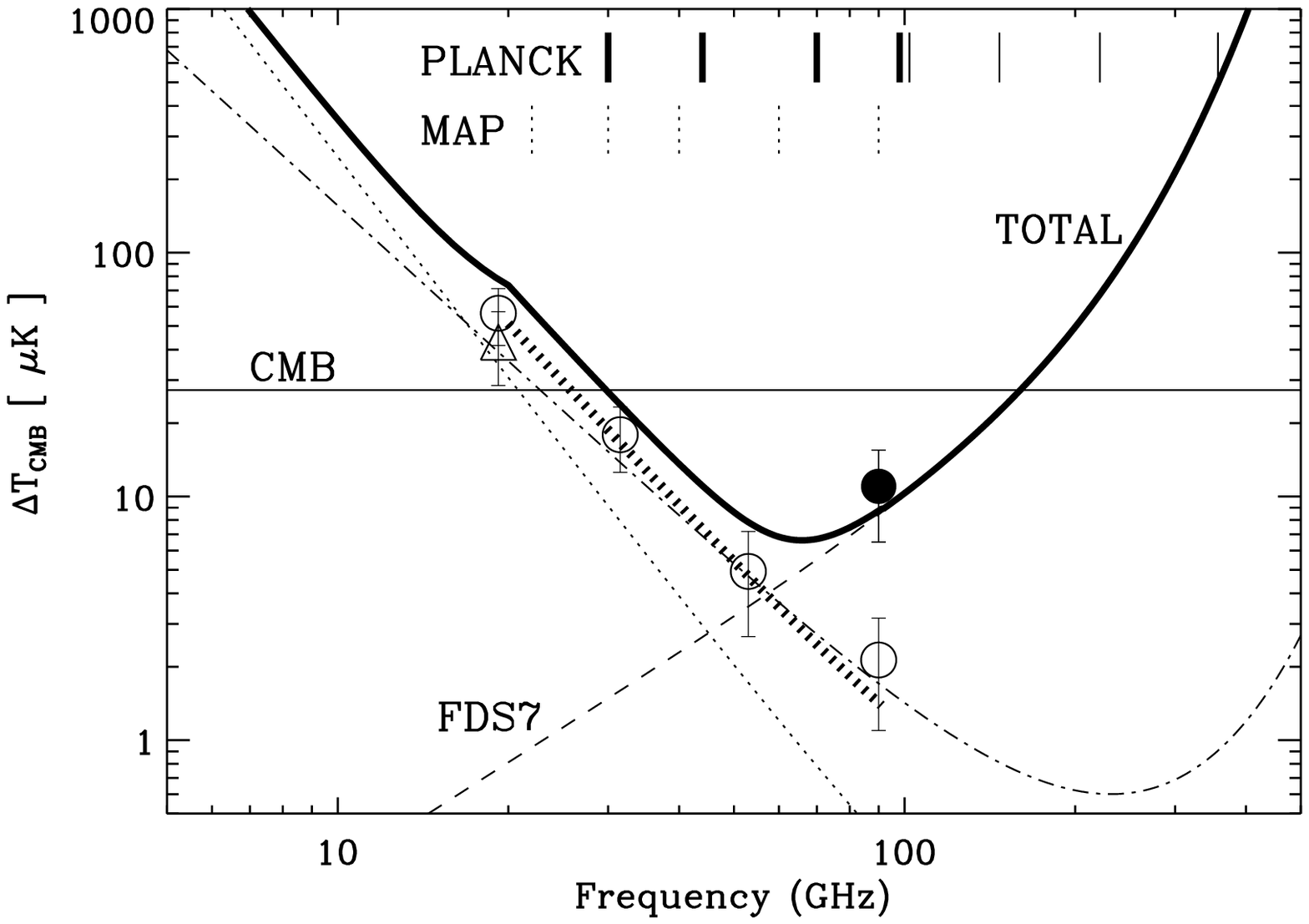}
  \caption{Foreground contaminations for the standard Galactic
  cut ($|b| \gtsim 20\degr$) at angular scales $\gtsim 7$\degr. Best-fitting models to 19.2~GHz and
  DMR data are shown for each
  foreground component: synchrotron (dotted line), free-free
  (dot-dashed line), thermal dust (dashed line) and anomalous dust
  (heavy dotted line between 19 and 90~GHz). Units have been converted
  to thermodynamic temperature fluctuations relative to $T_{\rm
  CMB}=2.73$~K; CMB fluctuations are represented as a horizontal line at a level of
  $\Delta T/T_{\rm CMB} =10^{-5}$ for comparison. Data from this paper
  are: synchrotron 19.2~GHz (open triangle), anomalous emission at
  31.5, 53 and 90~GHz (open circles) and total dust emission at 90~GHz
  (filled circle). The thick solid line gives an
  estimate total foreground level by adding all the foreground models in
  quadrature. The frequency channels for the {\it WMAP} (short vertical dashed lines) and
  {\it Planck} (short vertical solid lines) CMB satellites are shown at the top of
  the figure. The Low
  and High Frequency Instruments (LFI and HFI) for {\it Planck} are drawn as heavy and
  fine lines respectively. }
  \label{fig:foregrounds}
\end{figure}

Fig.~\ref{fig:foregrounds} summarises the integrated amplitudes of
Galactic foregrounds at large ($\gtsim 7$\degr) angular scales over the most
important frequencies for CMB studies. Our best models for
synchrotron ($\beta=3.02$ scaled to the 408~MHz map), free-free (scaled \ha\ predictions assuming $T_{e}=7000$
K; dot-dashed line) and
thermal dust model (FDS7; dashed line) emissions are plotted in terms of thermodynamic temperature
units (relative to $T_{\rm CMB}=2.73$~K). The best-fit anomalous component is
shown (heavy dotted line) over the range $19-90$~GHz. Data points are
shown for synchrotron at 19.2~GHz (open triangle) anomalous dust at 19.2, 31.5, 53 and 90~GHz (open circles),
total dust at 90~GHz (filled circle). The CMB signal is plotted at a
level of $\Delta T/T=10^{-5}$~K. The approximate total foreground level is
represented by adding the four components in quadrature, with the
anomalous component set to 52 \uk\ (the value at 19.2~GHz) for
frequencies below 19.2~GHz. This suggests that the minimum foreground
contamination for these data, is $\sim 70$~GHz. This is in good
agreement with the {\it WMAP} analysis which predicts the minimum to lie
at $\sim 60$ GHz.

The amplitudes for each of the foreground signals are now relatively well
known as shown in Fig. \ref{fig:foregrounds}. Strictly, Fig. \ref{fig:foregrounds} applies only for angular scales
of $7\degr$ and larger, and for all-sky observations with a Galactic cut
of $|b| \gtsim 20\degr$. For different Galactic cuts, there are likely to
be variations of a factor of $2-4$ for reasonable Galactic cuts as indicated
in Table 1. Observations covering smaller regions of sky, perhaps cleaner
regions (for example, away from the North Polar Spur), may be have reduced
foreground contamination by a factor of $\sim 2$. Finally, at higher
$\ell$-values (smaller angular scales), it is known that diffuse
Galactic foregrounds reduce in amplitude. The power spectra of Galactic
emission have a typical power-law indices ($C_{\ell} \propto
\ell^{-\alpha}$) of $\sim 2-3$ (Giardino et al. 2001). Thus at $1\degr$ angular scales,
the Galactic foregrounds, in terms of $\Delta T$, will be reduced by a factor $\sim 2$. However,
at angular scales $\ltsim 1$\degr, discrete extragalactic sources become the dominant foreground.

With relevance to the satellite missions {\it WMAP} and {\it Planck}, our
analysis implies that the lowest level of foreground contamination from diffuse
Galactic foregrounds at least on a 7\degr angular scale will be
between 50 and 100~GHz. 
This has been confirmed by Bennett et al (2003b) and seems to hold to
yet smaller scales $\sim$ 1\degr.
Our study also predicts that for the {\it WMAP} experiment,  the predominant
foreground at 22, 30 and perhaps 40~GHz, will be due to the
dust-correlated component. This is indeed found to be the case
by the recent {\it WMAP} analysis, although the interpretation
of the physical origin of the emission remains open to debate.
More accurate determination of its spectral behaviour, allied to lower
frequency observations (5-15 GHz) and higher angular resolution observations,
should allow the mechanism behind this unexpected foreground
to be ascertained. 
Moreover, given the likely region of minimal
foreground contamination in frequency space, the importance of the 
{\it Planck}-LFI instrument to the {\it Planck} mission as a whole cannot be overemphasised.

\subsection*{Acknowledgments}

The $COBE$ datasets were developed by the NASA Goddard Space Flight 
Center under the guidance of the $COBE$ Science Working Group and 
were provided by the NSSDC. The Wisconsin H-Alpha Mapper project is funded 
by the National Science Foundation. The Southern H-Alpha Sky Survey
Atlas project is supported by the NSF. We acknowledge Dave Cottingham for
providing us with a copy of the 19.2~GHz survey.


\bibliographystyle{mnras}

\label{lastpage}

\end{document}